\documentclass[english,aip,letter,superscriptaddress,twocolumn,floatfix,jcp,nofootinbib]{revtex4}
\usepackage[utf8]{inputenc}
\usepackage[T1]{fontenc}
\usepackage{graphicx}
\usepackage{amsmath,amssymb}
\usepackage{chemformula}
\usepackage[hidelinks]{hyperref}

\newcommand{\llangle}{\left\langle}
\newcommand{\rrangle}{\right\rangle}

\newcommand{\esc}{\!\cdot\!}

\begin{document}

\title{Unraveling Internal Friction in a Coarse-Grained Protein Model}

\author{Carlos Monago}
\affiliation{Dept.  F\'{\i}sica  Fundamental, Universidad Nacional
  de   Educaci\'on   a   Distancia, Madrid 28015,  Spain}
\author{J.A. de la  Torre}
\affiliation{Dept.  F\'{\i}sica  Fundamental, Universidad Nacional
  de   Educaci\'on   a   Distancia, Madrid 28015,  Spain}
\author{R. Delgado-Buscalioni}
\affiliation{Dept.   F\'{\i}sica   de  la   Materia  Condensada,
  Universidad Autónoma de Madrid, Madrid 28049, Spain}
\author{ Pep Espa\~nol}
\email{pep@fisfun.uned.es}
\affiliation{Dept.  F\'{\i}sica  Fundamental, Universidad Nacional
  de   Educaci\'on   a   Distancia, Madrid 28015,  Spain}

\date{}

\begin{abstract}
  Understanding the dynamic behavior  of complex biomolecules requires
  simplified models that not only  make computations feasible but also
  reveal fundamental mechanisms. Coarse-graining (CG) achieves this by
  grouping atoms into beads, whose  stochastic dynamics can be derived
  using  the Mori-Zwanzig  formalism,  capturing  both reversible  and
  irreversible  interactions.  In  liquid,  the dissipative  bead-bead
  interactions have so far  been restricted to hydrodynamic couplings.
  However, friction does not only arises from the solvent but notably,
  from the internal degrees of freedom  missing in the CG beads.  This
  leads  to an  additional  ``internal friction''  whose relevance  is
  studied in this contribution.   By comparing with all-atom molecular
  dynamics (MD), we neatly show  that in order to accurately reproduce
  the dynamics of  a globular protein in water  using a coarse-grained
  (CG) model,  not only a  precise determination of  elastic couplings
  and    the    Stokesian    self-friction    of    each    bead    is
  required.  Critically, the  inclusion of  internal friction  between
  beads is  also necessary  for a  faithful representation  of protein
  dynamics.  We  propose to  optimize the parameters  of the  CG model
  through a self-averaging method that integrates the CG dynamics with
  an evolution equation for the  CG parameters.  This approach ensures
  that selected  quantities, such as the  radial distribution function
  and the time correlation of bead velocities, match the corresponding
  MD values.
\end{abstract}

\pacs{}

\maketitle

\noindent{\small 
This article may be downloaded for personal use only. Any other use requires
prior permission of the authors and AIP Publishing. This article appeared in
C.~Monago, J.~A. de~la~Torre, R.~Delgado-Buscalioni, and P.~Espa\~nol,
J.~Chem.\ Phys.\ \textbf{162}, 114115 (2025) and may be found at
\url{https://doi.org/10.1063/5.0255498}.}

\section{Introduction}

Proteins dynamics  spans over  a broad range  of time  and length
  scales from fast aminoacid  vibrations \cite{2023RamanCG} over tens
of femtoseconds, to hundred  of nanoseconds required for ATP-activated
structural   changes,   leading  to   hydrodynamic-driven   collective
diffusion over longer time scales \cite{kapral2015}. In the picosecond
range,  low frequency  collective protein  vibrations are  critical to
enzyme function, promoting catalysis in ways whose microscopic details
are  still   largely  unknown   \cite{Cheatum2020,TOUSIGNANT2004}.   A
battery  of  experimental  techniques  is  used  to  sample  different
spatio-temporal domains \cite{grimaldo2019}.  Examples include various
sorts  of  Raman  spectroscopy  ($[0.1-100]  \mathrm{THz}$)  to  study
intermediate   and  fast   atomic  vibrations,   terahertz  near-field
microscopy  and neutron  scattering  \cite{2014Acbas}  to resolve  the
elusive  underdamped low  frequency  protein vibrations  out from  the
(solvent    and   peptide)    featureless   vibrational    background,
time-resolved X-ray diffraction in protein crystals \cite{Klyshko2024}
to sample  large (10-100 ns) conformational  changes, and fluorescence
techniques   \cite{schuler2018}    to   track   collective    diffusion   and
protein-protein   interactions    over   microns    and   microseconds
\cite{torrado2024}.   Experiments sample  coarse-grained variables  or
collective-variables  of the  macromolecule,  e.g.  polarizability  in
Raman  spectra  or  center  of mass  (CoM)  position  in  fluorescence
microscopy.  Sampling  often involves some external  oscillatory force
which  perturbs the  collective  variable  and in  this  case one  can
measure the force-response phase-lag to  access small fractions of the
forcing    period.     For     instance,    wall    oscillations    at
$[10-100] \text{MHz}$ introduced by quartz crystal microbalances (QCM)
measure  in-phase and  out-of-phase  shifts in  the  total wall  shear
stress, arising from the  presence of adsorbed macromolecules.  Atomic
force microscopy (AFM) in tapping  mode \cite{Phani2021} exerts 100 pN
forces at $\sim [10-100] \text{KHz}$  to sample the height of adsorbed
proteins and its interaction  with the substrate \cite{ZHANG20125141}.
There is  significant information hidden in  the viscoelastic response
of the  experimentally excited  collective variable.  The  central and
most  difficult   question  is   how  these  elastic   (in-phase)  and
dissipative  (out-of-phase) components  relate with  the biomolecule's
microscopic processes.  The  role of simulations is to  close this gap
between  experimental  signals  and   microscopic  details,  and  this
generally  requires an  intermediate  ``coarse  grained'' (CG)  level.
Following the  principle of Occam's  razor, the CG  description should
include only  the essential  details.  The  selection of  CG variables
represents the  first step in  establishing a simulation  protocol and
involves identifying the relevant length and time scales.  A recent CG
model has been proposed to reproduce surface enhanced Raman spectra by
grouping  a small  number of  atoms into  aminoacid-specific ``blobs''
\cite{2023RamanCG}.  By  contrast, a simple dumbbell  with few ``large
blobs'', can be used to reproduce the dipolar forces exerted by active
proteins  in hydrodynamic  media \cite{kapral2015}.   Indeed, proteins
have  been described  using  a  variable number  of  blobs (or  beads)
\cite{Noid2013,jin2022,Noid2023},  from the  $\alpha$  carbons of  the
protein            backbone           as            CG           sites
\cite{levitt1975,copperman2015,borges-araujo2023}  to  few  atoms  per
bead  as  in  the  MARTINI  framework  \cite{marrink2023},  to  larger
groupings          of         coherently          moving         atoms
\cite{amadei1993,tirion1996,kukol2008,Zhang2008,Zhang2009}.        The
decision on how  to group atoms is a subject  of considerable research
\cite{li2016,foley2020,giulini2020,jin2022,kidder2024}.    It  is
  clear that  the choice  of CG  variables should  be tailored  to the
  specific physical phenomena under investigation.

While significant  effort has  been dedicated to  accurately capturing
static equilibrium structural properties, much less attention has been
directed toward dynamic properties, which are crucial for interpreting
experimental signals  under non-equilibrium conditions.

In  turn,
  understanding the dynamics of CG  variables requires a discussion of
  dissipation and  its origins.    Interactions  between biomolecules
and the  experimental probe  are strongly affected  by the  solvent or
more generally the hydrodynamic  environment (e.g.  in fluid membranes
\cite{panzuela2018solvent}).  Solvent  interaction substantially alter
the  experimental  viscoelastic  response,  as clearly  shown  in  QCM
\cite{BuscalioniSM21,QCM24}  and AFM  \cite{Phani2021,Ruggeri2019} and
even in GHz-THz  spectroscopic techniques \cite{Cheatum2020}.  Lacking
a connection with the underlying levels, in experiments elasticity and
friction or  ``dissipation'' are generally introduced  and interpreted
in phenomenological intuitive ways.  The role of the solvent crucially
depends  on  the length  scale:  from  a  soft molecular  scaffold  at
sub-nanometer  range  \cite{Tan2019}  to  a  fluctuating  hydrodynamic
environment at longer scales.  Modelling the hydrodynamic scales might
require solving  fast vorticity diffusion  in QCM \cite{QCM24}  or the
Stokes-limit   diffusion    \cite{BalboaUsabiaga2013};   either   just
including      self-diffusion     or      collective     hydrodynamics
\cite{kapral2015}.  In  protein and  polymer dynamics  models, solvent
hydrodynamics  has   been  successfully  introduced   using  different
techniques              (based              on              Lagrangian
\cite{okuwaki2020,vaiwala2021,wang2024},                      Eulerian
\cite{peskin2002,Usabiaga2013,Delong2014}           or           using
Green-function-based               Brownian              hydrodynamics
\cite{carrasco1999,geyer2011,copperman2015,sedeh2018,tworek2023}).

In CG  models, dissipation
  occurs not only due to friction  with the solvent (wet friction) but
  also from the elimination of atomic degrees of freedom in favor of a
  bead  representation (dry  or internal  friction).  This  additional
  dissipation has received significant interest from both experimental
  \cite{ansari1992,borgia2012,soranno2012,soranno2018,schuler2018,Das2022}
  and                    computational                    perspectives
  \cite{echeverria2014,deSancho2014,zheng2015,daldrop2018,Dalton2023,Dalton2024}.
  Experimentally, the two sources  of dissipation can be distinguished
  by varying  the solvent's viscosity  — since this  parameter affects
  only the wet friction. It is  important to note that dissipation, as
  measured by entropy production, depends  on the level of description
  in much  the same way  that entropy does. Therefore,  dissipation is
  meaningful only within  the framework of the specific  CG model used
  to  interpret the  experiment,  and different  models  will lead  to
  different  ``internal frictions''.   In the  study of  intrinsically
  disordered and  unfolded proteins,  CG dynamics are  often described
  using the Rouse model with internal friction (RIF) \cite{cheng2013},
  where the two types of friction are represented.  In particular, the
  internal frictional  force is proportional to  the relative velocity
  between  neighboring   beads  along  the  polymer   chain.   In  the
  description of folding proteins, a usual model is a one-dimensional
  diffusion along a reaction  coordinate, typically a distance between
  residues,  where  Kramers  theory  is used  to  interprete  results
  \cite{borgia2012}.  The  parameters quantifying ``friction'' in different models are
  not comparable,  as they  have different physical  dimensions. Thus,
  the  interpretation  of   friction  is  inherently  model-dependent,
  underscoring the  importance of careful model  selection in deriving
  meaningful insights into protein dynamics.

In   this    paper,   we    provide   a

microscopic   motivation   for   bead-based   models   for
biomolecules consistently including

internal  and solvent
  frictions.            Mori-Zwanzig            (MZ)           theory
\cite{Zwanzig1961,Mori1965,Grabert1982,Zwanzig2001,Kinjo2007,hijon2010a,Schilling2022,izvekov2013,izvekov2017a,izvekov2017b,izvekov2019,izvekov2021}  provides  a  rigorous ``bottom-up''  approach  to
  derive  CG dynamics  by  connecting the  model  parameters with  the
  underlying all-atom  (AA) microscopic  dynamics via the  free energy
  and Green-Kubo formulae.

This route has been successfully  used to describe star polymers melts
\cite{hijon2010a,Li2014}  and  simple  liquids  \cite{Han2018,han2021}
but,  surprisingly,  it  has  not  been  applied  to  distinguish  the
different  origins  of  friction  in  macromolecular  solutions.   The
challenge of using stochastic  differential equations for CG variables
derived from the Mori-Zwanzig approach lies in the dependence of their
building blocks on conditional expectations.  As a result, they become
complex  many-body  functions  of   the  CG  variables,  making  their
computation non-trivial.  One approach to address this challenge is to
introduce parameterized models that retain  the structural form of the
microscopically derived building blocks, which then require estimation
of  the model  parameters.   Here, we  estimate  the CG-parameters  by
enforcing  that  averages  and \emph{time  correlations}  of  selected
observables computed at  the CG level match  those previously obtained
from  the   all-atom  (AA)  system.    This  is  achieved   through  a
self-averaging method that couples the dynamics of the CG variables to
a simple  relaxation equation for  the CG parameters  themselves.  The
coupled  dynamics  reaches  a  fixed  point  at  the  target  matching
conditions,       assuming      the       system      is       ergodic
\cite{kiefer2001,givon2004}. This self-averaging  method for parameter
estimation can  be used in other  CG descriptions as, for  example, to
extract friction channels in  different scenarios, including vorticity
transport  or   even  memory  kernels  arising   from  solvent-protein
molecular effects.  Here  we focus on a simple case:  a single globular protein
in  water.  The  reversible part  describes elastic  and electrostatic
forces between  beads, while the irreversible  part describes friction
between beads  (internal friction)  and (self)  friction of  the beads
with the  surrounding water  solvent.  The resulting  model is  thus a
combination of Dissipative Particle Dynamics and Langevin dynamics for
the beads.  We show, by  comparing with all-atom simulations, that the
internal   friction   is   \emph{essential}  to   reproduce   velocity
time-correlation of the single beads, and to interpret the vibrational
properties of a single protein in water.

\section{Raspberry model for a protein}
\begin{figure}[h]

  \includegraphics[width=0.4\textwidth]{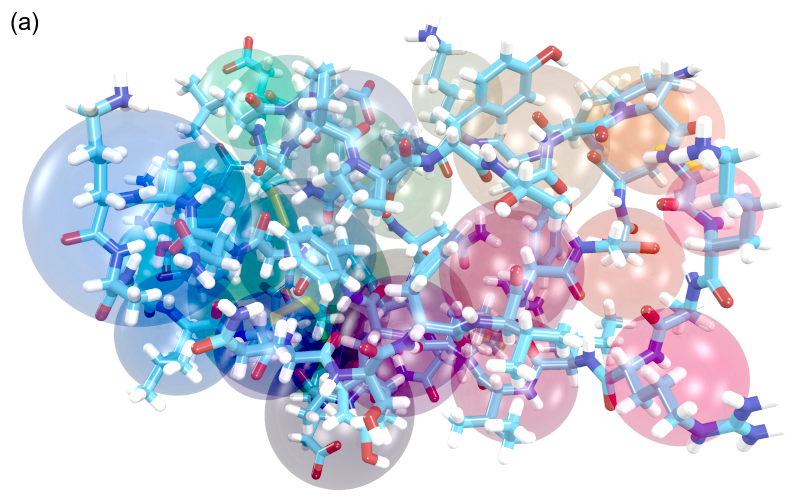}

  \includegraphics[width=0.4\textwidth]{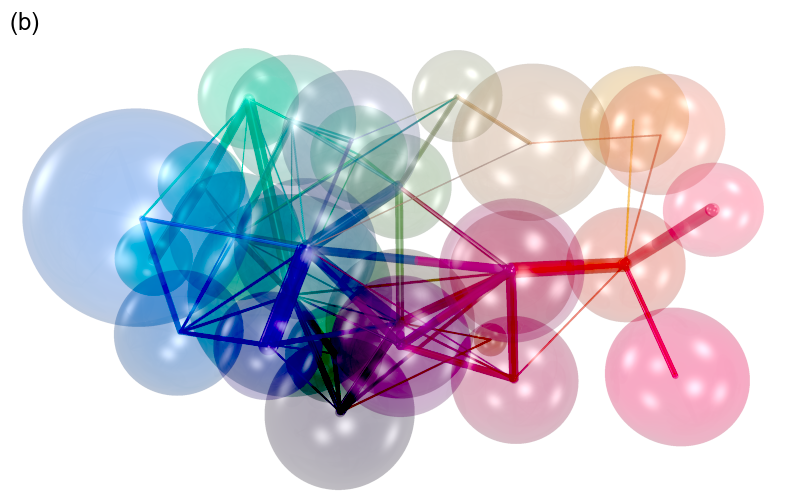}
  \caption{(a) The Bovine Pancreatic Trypsin Inhibitor protein (BPTI),
    displaying both the atomic structure  and the CG beads arranged to
    minimize   the  disparity   between   microscopic   and  CG   mass
    densities.  The radius  of each  bead is  scaled according  to its
    mass. (b) A representation of the elastic network, where the width
    of   each    link   corresponds    to   its    elastic   constant,
    $\kappa_{\mu\nu}$. Links with $\kappa_{\mu\nu}  < 10^{-3}$ are not
    shown. Water molecules not shown for clarity.}
  \label{Fig:1}

\end{figure}

We develop  a CG  ``raspberry'' model  of a  single protein  in water,
derived from first principles and based on a bead representation.  For
this  study,  we  focus  on the  globular  Bovine  Pancreatic  Trypsin
Inhibitor (BPTI) protein,  as shown in Fig  \ref{Fig:1}. The Molecular
Dynamics (MD) model is composed by a collection of atoms of mass $m_i$
with positions  ${\bf r}_i$ and  momenta ${\bf p}_i$.  We  denote with
$z$ the microstate of the full  system, that collects the position and
momenta  of  all particles,  including  water  molecules. Funtions  of
microstate     are    denoted     with    a     caret    $\hat{A}(z)$.

\subsection{The  CG  mapping} At  a  CG  level, the  protein  is
represented  as a  collection  of interacting  beads,  with each  bead
encompassing specific atoms of the protein.
The mesoscopic state
of the protein is defined by  coarse grained beads with center of mass
position   $\hat{\bf   R}_\mu$    and   momentum   $\hat{\bf   P}_\mu$
($\mu=1,\cdots,N_{\rm CG}$).  In this  work, $N_{\rm CG}=28$. These CG
variables are defined in the usual way
\begin{align}
  \label{eq:1}
  \hat{\bf R}_\mu(z)
  &=\sum_i\frac{m_i}{M_\mu}{\bf r}_i \delta_{\mu}(i)
    \nonumber\\
  \hat{\bf P}_\mu(z)
  &=\sum_i{\bf p}_i\delta_{\mu}(i)
\end{align}
The indicator function  $\delta_\mu(i)$ takes the value 1  if atom $i$
is   in  bead   $\mu$,  and   zero  otherwise   and  defines   the  CG
\textit{mapping}.   Many different  ways of  grouping atoms,  i.e.  of
specifying   $\delta_\mu(i)$,  in   a  protein   have  been   proposed
\cite{li2016,foley2020,giulini2020,jin2022,borges-araujo2023,kidder2024}. An elegant
  way to do this by using a Bayesian approach is given in \cite{chen2017a}. We follow a different strategy from these works.
As our  long term  interest is in  the elasto-mechanics  properties of
nanoscopic objects, the mapping that  we consider aims to minimize the
disparity  between  the mass  distribution  of  atoms and  beads,  and
resembles   a    soft-Voronoi   partition   of   atoms    into   beads
\cite{lu2005,zhang2010,tworek2023}. The  idea  is  to  minimize  the
difference  between  the microscopic  $  \hat{\rho}({\bf  r})$ and  CG
$ \overline{\rho}({\bf r})$ density fields, defined according to
\begin{align}
  \label{eq:2}
  \hat{\rho}({\bf r})&\equiv\sum_i^Nm_i\delta({\bf r}-{\bf r}_i),
\quad\quad
    \overline{\rho}({\bf r})\equiv\sum_\mu^MM_\mu \Delta({\bf r}-{\bf R}_\mu)
\end{align}
where $m_i$ is the  mass of atom $i$, $M_\mu$ is the  mass of the bead
$\mu$ and the Gaussian weight function is
\begin{align}
  \label{eq:4}
  \Delta({\bf r}-{\bf R}_\mu)=\frac{1}{Z}\exp\left\{-\frac{({\bf r}-{\bf R}_\mu)^2}{2\overline{\sigma}^2}\right\}
\end{align}
where  $Z$ ensures  that  the  Gaussian is  normalized  to unity,  and
$\overline{\sigma}$ is the  width of the Gaussian. By  dividing by the
total  mass of  the  protein,  we may  interpret  $\hat{\rho}({\bf r})$  and
$\overline{\rho}({\bf   r})$   as  probability   distributions   whose
discrepancy  can  be  minimized   through  the  Kullback-Leibler  (KL)
divergence, defined as
\begin{align}
  \label{KL}
  D(\rho|\overline{\rho})&=\int d{\bf r}\hat{\rho}({\bf r})\ln\frac{\hat{\rho}({\bf r})}{\overline{\rho}({\bf r})}
\end{align}
The KL  divergence becomes  a function of  the bead  positions through
$\overline{\rho}({\bf  r})$, and  we  ask which  configuration of  the
beads  gives  the  smallest  KL  divergence.  Therefore,  we  minimize
(\ref{KL})
\begin{align}
  \label{eq:15}
  \frac{\partial   D(\rho|\overline{\rho})}{\partial {\bf R}_\mu}
  &=-\int d{\bf r}\hat{\rho}({\bf r})\frac{\partial}{\partial {\bf R}_\mu}\ln\overline{\rho}({\bf r})
\end{align}
where   the   $\hat{\rho}\ln\hat{\rho}$   term,    being   independent  on   $R$,
dissapears.  Setting  to  zero   the  partial  derivatives  and  using
(\ref{eq:2}) for the CG density field gives the following condition
\begin{align}
  \label{eq:17}
  0=\int d{\bf r}\frac{\hat{\rho}({\bf r})}{\overline{\rho}({\bf r})}\exp\left\{-\frac{({\bf r}-{\bf R}_\mu)^2}{2\overline{\sigma}^2}\right\}({\bf r}-{\bf R}_\mu)
\end{align}
Inserting the form (\ref{eq:1}) for the microscopic density field, (\ref{eq:17}) becomes
\begin{align}
    \label{eq:36}
\sum_im_i     \frac{1}{\overline{\rho}({\bf r}_i)}\exp\left\{-\frac{({\bf r}_i-{\bf R}_\mu)^2}{2\overline{\sigma}^2}\right\}({\bf r}_i-{\bf R}_\mu)=0
\end{align}
that can be written as
\begin{align}
  \label{eq:37}
  {\bf R}_\mu&=\frac{\sum_im_i{\bf r}_i\chi_\mu({\bf r}_i)}{\sum_im_i\chi_\mu({\bf r}_i)}
\end{align}
where we have introduced the Shepard functions
\begin{align}
  \label{eq:38}
  \chi_\mu({\bf r})&=\frac{\exp\left\{-\frac{({\bf r}-{\bf R}_\mu)^2}{2\overline{\sigma}^2}\right\}}{\sum_\nu\exp\left\{-\frac{({\bf r}-{\bf R}_\nu)^2}{2\overline{\sigma}^2}\right\}}
\end{align}
that form a partition of unity
\begin{align}
  \label{eq:39}
  \sum_\mu\chi_\mu({\bf r})=1 ,\,\forall\,{\bf r}
\end{align}
These functions  depend on the  width of the  Gaussians and it  can be
shown  that  as  $\overline{\sigma}\to0$,  this  is,  for  very  sharp
Gaussians, the  Shepard functions tend to  the characteristic function
of the Voronoi cells centered around ${\bf R}_\mu$ \cite{Flekkoy1999}.

Equation  (\ref{eq:37}) is  a  non-linear equation  for ${\bf  R}_\mu$
because  the   Shepard  functions   depend  on  the   beads  positions
themselves. Therefore, the positions of  the beads will be obtained in
an iterative way, reminiscent of the Lloyd's algorithm for calculation
of centroidal tessellations \cite{lloyd1982}. This is, we denote ${\bf R}^n_\mu$ as the
solution at the $n$-th iteration and write the iteration as
\begin{align}
  \label{eq:40}
    {\bf R}^{n+1}_\mu&=\frac{\sum_im_i{\bf r}_i\chi^n_\mu({\bf r}_i)}{\sum_im_i\chi^n_\mu({\bf r}_i)}
\end{align}
where $\chi^n_\mu({\bf r})$ is the Shepard function defined in terms
of the bead's positions ${\bf  R}^n_\mu$ at the $n$-th iteration. Once
the iterative  procedure has converged  we have the best  positions of
beads  that would  reproduce  the microscopic  density  (at the  scale
$\overline{\sigma}$). After convergence, the  mapping appearing in the
definition   of   the  CG   density field (\ref{eq:2})  is   provided  by
$\delta_\mu(i)=\chi_\mu({\bf  r}_i) $.   This partition  is computed
with  the  initial configuration  and  remains  constant in  the  time
evolution (this is,  the particles constituting a bead  are always the
same). The selected value of  $\sigma=0.053$ is sufficiently small for
$\delta_\mu(i)$ to be very close to 0 or 1.

\subsection{The  CG dynamics}  The  objective of  the theory  of
coarse-graining  is  to  produce  the  equations  of  motion  for  the
mesoscopic  variables $\hat{\bf  R}_\mu,\hat{\bf P}_\mu$.   While
  this   was   achieved   by   the   pioneering   works   of   Zwanzig
  \cite{Zwanzig1961} and Mori \cite{Mori1965},  there has been renewed
  interest in  the field in  recent times and  a number of  works have
  addressed  the  construction of  CG  equations  from the  underlying
  microscopic                   Hamiltonian                   dynamics
  \cite{Grabert1982,Zwanzig2001,Kinjo2007,hijon2010a,Schilling2022,izvekov2013,izvekov2017a,izvekov2017b,izvekov2019,izvekov2021}.
We  follow  Ref.   \cite{hijon2010a},  where  we  derived  the  closed
non-linear  Stochastic Differential  Equation (SDE)  that governs  the
dynamics of the CG variables {in Eq.~}{(\ref{eq:1})} using the Zwanzig
projector method.  The  only assumption required to arrive  at the SDE
is that the bead variables evolve in two separated time scales, a slow
one  due to  the  large size  of  the beads,  and a  fast  one due  to
collisions  and  vibrations. This  enables  the  bead dynamics  to  be
described  as a  Markov  diffusion  process governed  by  a SDE.   One
condition that the beads should satisfy to comply with this assumption
is that they should be sufficiently massive, in order for the velocity
of the bead to  evolve in a time scale much larger  than the forces on
the bead.  We  have verified that this is reasonably  satisfied in our
description,  where the  force-force  decorrelation is  about 4  times
faster than  that of  the velocity  (see Supplementary  Material (SM),
Section S1).

The Ito SDE governing the beads are \cite{hijon2010a}
\begin{align}
  \label{SDE-RP}
  d{\bf R}_\mu
  &={\bf V}_\mu dt
    \nonumber\\
  d{\bf P}_\mu
  &=\llangle\hat{\bf F}_\mu\rrangle^{RP}dt
    -\sum_\nu \boldsymbol{\Gamma}_{\mu\nu}\esc {\bf V}_\nu dt
    +d\tilde{\bf P}_\mu
\end{align}
where  ${\bf  V}_\mu={\bf  P}_\mu/M_\mu$   is  the  velocity  of  bead
$\mu$. The conditional average is defined as
\begin{align}
  \label{eq:7}
  \llangle\cdots\rrangle^{RP}&=\int dz \rho^{\rm eq}(z)\prod_\mu^M\delta(\hat{\bf R}_\mu(z)-{\bf R}_\mu)
                               \delta(\hat{\bf P}_\mu(z)-{\bf P}_\mu)\cdots
\end{align}
where $\rho^{\rm eq}(z)$ is the equilibrium ensemble of the protein+water
system.  The state  dependent  friction tensor  is  defined through  a
Green-Kubo formula
\begin{align}
  \label{GK}
  \boldsymbol{\Gamma}_{\mu\nu}(R,P)
  &=\frac{1}{k_BT}\int_0^{\Delta t}dt \llangle \delta \hat{\bf F}_\mu(t)\delta \hat{\bf F}_\nu\rrangle^{RP}
\end{align}
where    the    fluctuation   of    the    force    is   defined    as
$  \delta   \hat{\bf  F}_\mu=   \hat{\bf  F}_\mu-   \llangle  \hat{\bf
  F}_\mu\rrangle^{RP}$.   The  random  noise $d\tilde{\bf  P}_\mu$  in
Eq.~{(\ref{SDE-RP})} satisfies the Fluctuation-Dissipation theorem
\begin{align}
  \label{FDT}
  d\tilde{\bf P}_\mu  d\tilde{\bf P}_\nu=2k_BT\boldsymbol{\Gamma}_{\mu\nu}dt
\end{align}
The  random noise  can be  expressed as  a linear  combination of  the
Wiener  process.  As  we will  see, the  inspiration to  construct the
particular  linear  combination  entering  the noise  comes  from  the
heuristic identification of the noise  $ d\tilde{\bf P}_\mu $ with the
fluctuating force $ \delta \hat{\bf F}_\mu$.

The   SDEs  {in   Eq.~}{(\ref{SDE-RP})}  are   equivalent  to   a
Fokker-Planck   Equation   that   governs   the   evolution   of   the
non-equilibrium probability density $P(R,P,t)$ towards the equilibrium
probability $P^{\rm  eq}(R,P)$.  By  definition, $P^{\rm  eq}(R,P)$ is
given by the pushforward
\begin{align}
  P^{\rm eq}(R,P)=\int dz \rho^{\rm eq}(z)
  \prod_\mu
  \delta(\hat{\bf R}_\mu(z)-{\bf R}_\mu)
  \delta(\hat{\bf P}_\mu(z)-{\bf P}_\mu)
\end{align}
In the canonical ensemble,
the Gaussian momentum integrals are easily performed, leading to
\begin{align}
  \label{eq:11}
  P^{\rm eq}(R,P)=\frac{1}{Z}\exp\left\{-\beta\left[\sum_\mu\frac{{\bf P}^2_\mu}{2M_\mu}+ V^{\rm MF}(R)\right]\right\}
\end{align}
where the potential of mean force (PMF) is defined by
\begin{align}
V^{\rm MF}(R)=-k_BT\ln\int dr e^{-\beta V(r)}
  \prod_\mu  \delta(\hat{\bf R}_\mu(z)-{\bf R}_\mu)
  \label{VMF}
\end{align}
and $Z$ renders  the probability given by Eq.~{(\ref{eq:11})} normalized.   It can be
shown  that  the  conditional  expectation of  the  microscopic  force
appearing in Eq.~{(\ref{SDE-RP})} can  be written as  the gradient  of this
potential of mean force \cite{hijon2010a}
\begin{align}
  \label{eq:12}
  \llangle \hat{\bf F}_\mu\rrangle^{RP}=-\frac{\partial V^{\rm MF}}{\partial {\bf R}_\mu}(R)
\end{align}

The  SDEs given  in  Eq.~{(\ref{SDE-RP})} are  very general:  Any
Markovian bead description in terms  of their positions and velocities
will have the same SDE.   However, different systems (star polymers as
in \cite{hijon2010a,Faure2017}  or proteins here) will  have different
functional forms for  the potential of mean force  $V^{\rm MF}(R)$ and
the friction  tensors $\boldsymbol{\Gamma}_{\mu\nu}(R)$.   In general,
these are complex many-body functions  of the system's macrostate $R$,
making  them  challenging  to  compute as  they  are  defined  through
conditional expectations. A practical approach is to approximate these
many-body functions using a sufficiently flexible parametric model.

\subsection{Model for the PMF}
For a bead representation of a  system of bonded particles as those in
a protein, we expect the behavior  of the beads to display an elastic
behavior.  A particularly  useful and simple model is  given by beads
connected with Frenkel  springs \cite{hirschmann2023}. In the course  of this investigation,
we have  realized that  this model  is not  sufficient to  capture the
dynamics of  the beads.  It turns  out that the beads  have net charge
and, therefore,  it is convenient  to introduce an  additional Coulomb
electrostatic interactions.   The potential of mean  force in Eq.~{(\ref{VMF})}
is \textit{modeled} with the following  parameterized  pair-wise model
\begin{align}
  \label{eq:47}
  V^{\rm MF}(R)=\frac{1}{2}\sum_{\mu\nu}^2\kappa_{\mu\nu} (R_{\mu\nu}-\overline{R}_{\mu\nu})^2+ \omega_{\mu\nu}C\frac{q_\mu q_\nu}{R_{\mu\nu}}
\end{align}
where          $R_{\mu\nu}=|{\bf           R}_{\mu\nu}|$          with
${\bf  R}_{\mu\nu}={\bf R}_\mu-{\bf  R}_\nu$ is  the distance  between
beads, $\overline{R}_{\mu\nu}=\llangle R_{\mu\nu}\rrangle^{\rm eq}$ is
the equilibrium distance, $\kappa_{\mu\nu}$ is the elastic constant of
the spring joining beads $\mu$ and $\nu$, $C$ is the Coulomb constant,
$q_\mu$ is  the net  charge of bead  $\mu$, and  $\omega_{\mu\nu}$ are
weights  that   modulate  the   electric  interaction   between  beads
accounting   for  possible   screening  effects   introduced  by   the
surrounding water.

\subsection{Model for the friction tensor}
The  velocity  dependent  friction  force in {Eq.~}{(\ref{SDE-RP})}
captures several physical  effects that can be inferred  from the very
structure  of  the Green-Kubo  formula  in Eq.~{(\ref{GK})}.   The total  force
$\hat{\bf    F}_\mu    $    on    $\mu$-th   bead    is    given    by
$\hat{\bf F}_\mu  =\hat{\bf F}^{\rm b}_{\mu}+\hat{\bf  F}^{\rm w}_\mu$
where $\hat{\bf F}^{\rm b}_{\mu}$ is the force due to the other beads,
and $\hat{\bf F}^{\rm w}_\mu$ is the force due to the water molecules.
The  force on  a bead  due to  the  rest of  beads can  be written  in
pair-wise                                                         form
$\hat{\bf F}^{\rm b}_{\mu}=\sum_\nu\hat{\bf F}^{\rm b}_{\mu\nu}$ where
$        \hat{\bf         F}^{\rm        b}_{\mu\nu}=\sum_{ij}\hat{\bf
  F}_{ij}\delta_{\mu}(i)\delta_\nu(j)$, and $\hat{\bf  F}_{ij}$ is the
force that the  $j$-th atom of the protein exerts  on the $i$-th atom.
Because  Newton's  Third   Law  $\hat{\bf  F}_{ij}=-\hat{\bf  F}_{ji}$
implies $\hat{\bf F}^{\rm  b}_{\mu\nu}=-\hat{\bf F}^{\rm b}_{\nu\mu}$,
we have that $ \sum_\mu\hat{\bf F}^{\rm b}_{\mu}=0$.

The decomposition of the force on bead  $\mu$ as the sum of the forces
due  to other  beads  and  to water  entails  a  decomposition of  the
friction tensor Eq.~{(\ref{GK})} of the following form

\begin{align}
  \label{eq:13}
  \boldsymbol{\Gamma}_{\mu\nu}
  &=  \boldsymbol{\Gamma}^{\rm bb}_{\mu\nu}+\boldsymbol{\Gamma}^{\rm bw}_{\mu\nu}+\boldsymbol{\Gamma}^{\rm wb}_{\mu\nu}+\boldsymbol{\Gamma}^{\rm ww}_{\mu\nu}
\end{align}
where we have introduced the different friction tensors
\begin{align}
  \label{eq:14}
  \boldsymbol{\Gamma}^{\rm \alpha \beta }_{\mu\nu}
   &\equiv\frac{1}{k_BT}\int_0^{\Delta t}dt \llangle
     \delta \hat{\bf F}^{ \alpha}_{\mu}(t)\delta \hat{\bf F}^{\beta}_{\nu}
 \rrangle^{RP}
\end{align}
where $\alpha,\beta={\rm b, w}$.

These  four contributions  reflect  the different  types of  frictions
between         beads.          The         bead-bead         friction
$ \boldsymbol{\Gamma}^{\rm bb}_{\mu\nu}$  accounts for the dissipative
process induced by  the elimination of the protein  degrees of freedom
in favor  of the beads. Because  of Newton's third law  we
have that
\begin{align}
  \label{N3b}
\sum_\nu \boldsymbol{\Gamma}^{\rm bb}_{\mu\nu}  =0
\end{align}
The bead-bead friction tensor can be written
\begin{align}
\boldsymbol{\Gamma}^{\rm bb}_{\mu\nu}
  &=\sum_{\mu'\nu'}\frac{1}{k_BT}\int_0^{\Delta t}dt \llangle
    \delta \hat{\bf F}^{\rm b}_{\mu\mu'}(t)\delta \hat{\bf F}^{\rm b}_{\nu\nu'}
\rrangle^{RP}
\end{align}
We expect that the correlation of the force between different pairs of
beads    will   be    much    smaller    than   the    autocorrelation
\cite{espanol2016}.   Therefore,  for   $\mu\neq\nu$  we  assume
$\llangle \delta  \hat{\bf F}^{\rm  b}_{\mu\mu'}(t)\delta \hat{\bf
  F}^{\rm  b}_{\nu\nu'}\rrangle^{RP}\simeq0$, except  if $\mu'=\nu$
and $\nu'=\mu$, implying
\begin{align}
  \label{GK-b}
\boldsymbol{\Gamma}^{\rm bb}_{\mu\nu}
  &=\frac{1}{k_BT}\int_0^{\Delta t}dt \llangle
    \delta \hat{\bf F}^{\rm b}_{\mu\nu}(t)\delta \hat{\bf F}^{\rm b}_{\nu\mu}
\rrangle^{RP}
    \nonumber\\
  &=-\frac{1}{k_BT}\int_0^{\Delta t}dt \llangle
    \delta \hat{\bf F}^{\rm b}_{\mu\nu}(t)\delta \hat{\bf F}^{\rm b}_{\mu\nu}
    \rrangle^{RP}
\equiv-\boldsymbol{\gamma}_{\mu\nu}^{\rm bb}
\end{align}
where we  have used Newton's Third  Law and the last  line defines the
bead-bead friction  tensor $\boldsymbol{\gamma}_{\mu\nu}^{\rm  bb}$ in
terms  of  an  autocorrelation   function.   Finally,  on  account  of
Eq.~{(\ref{N3b})}, we write for all $\mu,\nu$
\begin{align}
  \label{Gbb}
  \boldsymbol{\Gamma}^{\rm bb}_{\mu\nu}
  &=\delta_{\mu\nu}\sum_{\mu'}\boldsymbol{\gamma}^{\rm bb}_{\mu\mu'}
    -\boldsymbol{\gamma}^{\rm bb}_{\mu\nu}
\end{align}
under  the convention  that the  pair friction  vanishes if  the two
beads  coincide,  $\boldsymbol{\gamma}^{\rm bb}_{\mu\mu}=0$.

The   Green-Kubo  expressions   are  many-body   functions  given   by
conditional expectations that cannot easily be computed, and judicious
modelization is required.   In principle, we could  use similar models
for  $\boldsymbol{\gamma}_{\mu\nu}^{\rm bb}$  as  those  used in  Ref.
\cite{hijon2010a} that depend  on both the distance  between the beads
and the  direction of the unit  vector joining them.  However,  in the
protein case,  the beads  are  bonded  and  they remain  at  certain
distances from the rest. Therefore, we choose the simpler model
\begin{align}
  \label{gbb}
  \boldsymbol{\gamma}^{\rm bb}_{\mu\nu}
  &=\gamma_{\mu\nu}^{\rm bb\bot}
    \left[{\bf 1}-{\bf e}_{\mu\nu}\esc {\bf e}_{\mu\nu}^T\right]
    +\gamma_{\mu\nu}^{\rm bb||}{\bf e}_{\mu\nu}\esc {\bf e}_{\mu\nu}^T
  &&\mu\neq\nu\end{align}
where
$\gamma_{\mu\nu}^{\rm bb\bot},\gamma_{\mu\nu}^{\rm bb||}$ are constant
friction  coefficients, \textit{independent}  on position.  Therefore,
the only position dependence in this model arises from the orientation
of    the    pair through the unit vector ${\bf e}_{\mu\nu}={\bf R}_{\mu\nu}/|{\bf R}_{\mu\nu}|$.
Observe that the model  in Eq.~{(\ref{gbb})} requires two
friction coefficients for  every pair, resulting in a  large number of
parameters.    We  will   simplify  matters   by  assuming   that  the
longitudinal and  tangential friction  coefficients are all  equal for
all                  pairs,                   this                  is
$\gamma_{\mu\nu}^{\rm    bb\bot}=\gamma^{\bot},   \gamma_{\mu\nu}^{\rm
  bb||}=\gamma^{||}$. We have checked that this simpler model yields
practically identical  results  than  the   more  complex  model  using  different
frictions for each pair.

Next, consider $  \boldsymbol{\Gamma}^{\rm ww}_{\mu\nu}$ that involves
the  correlation of  the force  $\hat{\bf F}^{\rm  w}_{\mu}$ that  the
water exerts on  bead $\mu$ and the force  $\hat{\bf F}^{\rm w}_{\nu}$
that the water exerts on bead $\nu$. When $\mu=\nu$ the friction force
on the blob $\mu$ is proportional to  the velocity of the blob, with a
coefficient  $\gamma^S$  that  we   interpret  as  a  Stokes  friction
coefficient.   For  $\mu\neq\nu$  the  forces are  correlated  due  to
hydrodynamic   interactions.     Therefore,   the    friction   tensor
$ \boldsymbol{\Gamma}^{\rm ww}_{\mu\nu}$ accounts for the hydrodynamic
interactions between  beads.  A  common strategy  in the  treatment of
water-mediated friction  between beads consists on  assuming the beads
are  spherical  particles to  which  the  results of  time-independent
linearized hydrodynamics  apply.  In this way,  beads interact through
the                    Rotner-Prager-Yamakawa                   tensor
\cite{kimMicrohydrodynamicsPrinciplesSelected1991}   where    even   a
distinction between ``internal'' and ``surface''  beads can be made to
improve results \cite{carrasco1999}. Nonetheless, applying the Green's
function of unbounded fluids appears somewhat incongruent for modeling
a compact,  relatively small  cluster of densely  packed beads  of the
kind  shown in  Fig  \ref{Fig:1}.   In addition, and estimate of the  time scales  of
propagation    of    hydrodynamic    interactions    gives
$a^2/\nu\simeq 19.5$ ps where $\nu\simeq 3.21\cdot 10^{-2}$ \AA$^2$/fs
is   the  kinematic   viscosity   of  water   for   the  TIP3P   model
\cite{gonzalez2010shear} and $a\simeq 25$  \AA$\,$ a typical dimension
of the protein.  Therefore, the time scale of propagation of continuum
hydrodynamic interactions is  two orders of magnitude  larger than the
typical time  scale of  the velocity  autocorrelation function  of the
beads,  which are  in  the  hundred of  femtoseconds  range (see  Fig.
\ref{Fig:2c}). In the time scale in which hydrodynamic interactions
would be effective,  the beads have already  been equilibrated through
the  self-friction with  water. In  this work,  we neglect  long range
hydrodynamic  interactions  between  beads, and  all  the  dissipative
effects of the solvent are captured with a simple Stokes friction.

Finally,  let us  consider the  cross correlations  between the  force
$\hat{\bf F}^{\rm  b}_{\mu\nu}$ that bead  $\nu$ exerts on  bead $\mu$
and  the force  $\hat{\bf  F}^{\rm  w}_{\mu'}$ of  the  water on  bead
$\mu'$. The pair-force $\hat{\bf F}^{\rm b}_{\mu\nu}$ between beads is
due to the  elastic forces within the protein, and  we expect that the
value of this force  has little to do with the force  due to the water
on another bead $\mu'$, which depends on the state of the water around
bead $\mu'$. Even in the case that $\mu'=\mu$ or $\mu'=\nu$, we expect
that the force  on the bead due  to the water or to  another bead will
have not strong correlation. Therefore,  we will neglect the effect of
the friction forces represented by these cross correlations and assume
$\boldsymbol{\Gamma}^{\rm        bw}_{\mu\nu}=\boldsymbol{\Gamma}^{\rm
  wb}_{\mu\nu}\simeq 0$.

\subsection{Final form of the parameterized bead model}
The SDE  Eq.~{(\ref{SDE-RP})} for  the positions and  momenta of  the beads
take  the  following   form  once  we  include   the  above  modeling
assumptions
\begin{align}
  \label{SDE-Model}  d{\bf  R}_\mu
  &={\bf  V}_\mu  dt  \nonumber\\
  d{\bf    P}_\mu
  &=   \left({\bf    F}^{\rm   rev}_{\mu}+{\bf    F}^{\rm irr}_{\mu}+{\bf F}^{\rm ran}_{\mu}\right)dt
\end{align}
The         reversible        force         is        given         by
${\bf F}^{\rm rev}_{\mu}\equiv \llangle \hat{\bf F}_\mu\rrangle^{RP}$.
From Eq.~{(\ref{eq:12})} and Eq.~{(\ref{eq:47})} this is
\begin{align}
  \label{Frev}
  {\bf        F}^{\rm       rev}_{\mu}
  &=
    -\sum_\nu\kappa_{\mu\nu}\left(R_{\mu\nu}-\overline{R}_{\mu\nu}\right){\bf
    e}_{\mu\nu}            +             \omega_{\mu\nu}C\frac{q_\mu
    q_\nu}{R_{\mu\nu}^2}{\bf e}_{\mu\nu}
\end{align}
The irreversible force is given by
\begin{align}
  \label{Firr}
  {\bf F}^{\rm irr}_{\mu}
  &= - \sum_\nu \left(\gamma^{\bot}\left[{\bf 1}-{\bf e}_{\mu\nu}\esc {\bf e}_{\mu\nu}^T\right]
    +\gamma^{||}{\bf e}_{\mu\nu}\esc {\bf e}_{\mu\nu}^T\right)\esc {\bf V}_{\mu\nu }
    -\gamma^S{\bf V}_\mu
\end{align}
The first contribution  gives a friction proportional  to the relative
velocities ${\bf V}_{\mu\nu}={\bf V}_\mu-{\bf V}_\nu$ of the beads, as
appears    in     the    Dissipative    Particle     Dynamics (DPD)   model
\cite{Espanol1995epl,espanol2017,Ellero2018},          and
represents  internal friction.   It  has  parallel and  perpendicular
components to the line joining  the beads.  The second contribution is
a Stokes friction force proportional to the velocity of the bead, with
$\gamma^S$ the friction coefficient, typical for the Langevin model of
Brownian motion.

Finally,   the    noise   terms   are   tailored    to   satisfy   the
Fluctuation-Dissipation   theorem.    The   random   force   has   two
contributions        due        to       beads        and        water
$ {\bf  F}^{\rm ran}_{\mu}dt=d\tilde{\bf  P}^{\rm b}_{\mu}+d\tilde{\bf
  P}^{\rm w}_{\mu}$.  The random force $d\tilde{\bf P}^{\rm b}_\mu$ is
a  white noise  approximation  to the  microscopically defined  random
force  $\delta\hat{\bf F}_{\mu}^{\rm  b}$ in  the Green-Kubo  friction given by Eq.~
{(\ref{GK-b})}. The microscopic force due to  the other beads is the sum
of      pair      forces       and,      therefore,      we      model
$   d\tilde{\bf   P}^{\rm   b}_\mu   =\sum_\nu   d\tilde{\bf   P}^{\rm
  b}_{\mu\nu}$,  where   $d\tilde{\bf  P}^{\rm  b}_{\mu\nu}$   is  the
stochastic noise representing the force that bead $\nu$ exerts on bead
$\mu$.   This  force  satisfies  Newton's Third  Law  and,  therefore,
$d\tilde{\bf  P}^{\rm  b}_{\mu\nu}=-d\tilde{\bf P}^{\rm  b}_{\nu\mu}$.
The structure of this force can be represented with a DPD random force
of the form \cite{espanol2016}
\begin{align}
  d\tilde{\bf P}^{\rm b}_{\mu\nu}
  &=\sqrt{2k_BT\gamma^{||}}
    {\bf e}_{\mu\nu} dW_{\mu\nu}
    +\sqrt{2k_BT\gamma^{\bot}}{\bf e}_{\mu\nu}\times d{\bf V}_{\mu\nu} \label{dPbrown}
\end{align}
where the independent increments of the Wiener processes satisfy
  \begin{align}
    dW_{\mu\mu'}    dW_{\nu\nu'}
    &=\left[\delta_{\mu\nu}\delta_{\mu'\nu'}+\delta_{\mu\nu'}\delta_{\nu\mu'}\right]dt
      \nonumber\\
    d{\bf V}^\alpha_{\mu\mu'}    d{\bf V}^\beta_{\nu\nu'}
    &=\left[\delta_{\mu\nu}\delta_{\mu'\nu'}+\delta_{\mu\nu'}\delta_{\nu\mu'}\right]dt\delta^{\alpha\beta}
      \nonumber\\
    d{W}_{\mu\mu'} d{\bf V}^\beta_{\nu\nu'} &=0
  \end{align}

  The random force describing the bombardment  of water on the bead is
  proportional  to  an independent  increment  of  the Wiener  process
  $ d{\bf U}_\nu$ as
  \begin{align}
  d\tilde{\bf P}^{\rm w}_{\mu}
        &=\sqrt{2k_BT\gamma^S}          d{\bf U}_\nu
          \label{Fin:Noises}
\end{align}
Details on the fulfillment  of the Fluctuation-Dissipation theorem are
provided in the SM Section~S2.

In  summary,  the  present microscopically  inspired  mesoscopic
model for the  dynamics of the beads assumes an  elastic network model
together with  Coulomb interactions for  the conservative part  of the
dynamics,  and  a combination  of  DPD  and  Langevin forces  for  the
irreversible  dynamics.  The  DPD  forces represent  \textit{internal}
dissipation due  to the CG  of the  protein into beads.   The Langevin
forces  account for  the interaction  of  the protein  beads with  the
surrounding water.

\section{Methods}
\label{Sec:REM}\subsection{MD simulations}
We have conducted  all atom (AA) MD simulations of  a protein in water
in  the NVE  ensemble.  Using  the NPT  ensemble, suitable  for static
equilibrium  properties, would  kill all  the dynamic  effects we  are
interested  in.  The  selected  protein is  Bovine Pancreatic  trypsin
inhibitor (BPTI), whose structure was taken from the Protein Data Bank
(PDB) \cite{wwpdbconsortium2019} with id 4PTI.  For the AA simulations
we used the force  field AMBER99SB-ILDN \cite{lindorff-larsen2010} for
the protein  and the  TIP3P water  model \cite{jorgensen1983}  for the
solvent.   The AA  simulations  were performed  using the  Large-scale
Atomic/Molecular  Massively   Parallel  Simulator   (LAMMPS)  software
\cite{thompson2022}.  We used  the SHAKE algorithm \cite{ryckaert1977}
to  constrain the  bonds  with hydrogen  atoms thus  allowing  a 2  fs
timestep.  For the Lennard-Jones  and Coulomb interactions the cut-off
was set  at 15  \AA{} and  long-range electrostatics  between all-atom
particles  were calculated  with  the particle–particle  particle-mesh
algorithm  \cite{hockney2021}.  The  AA simulations  were equilibrated
for 1 ns in the NVT ensemble at 300 K with a heat bath for the protein
and another for the solvent, then  we removed the bath associated with
the  protein  and we  let  the  system relax  for  4  ns.  Lastly,  we
equilibrate for 10 ps  in the NVE ensemble using a  0.2 fs timestep to
avoid unexpected behaviors due to  the removal of the thermostat.  The
sampling frequency for both the  equilibration and production runs was
set at 10  fs.  Unlike nearly all  molecular dynamics simulations
  of proteins  investigating internal  friction — which  are typically
  performed in the NVT \cite{daldrop2018,Dalton2023,Dalton2024} or NPT
  \cite{echeverria2014,deSancho2014,zheng2015}    ensembles   —    our
  production runs are  conducted in the NVE  ensemble. This procedure
ensures that no  artifacts are created in our  production runs.  Since
no thermostats are used during the production run, the actual dynamics
exhibits hydrodynamic behavior.  This allows  us to observe the effect
of  sound wave  propagation  in the  autocorrelation  function of  the
protein's center of  mass velocity.  This effect manifest  as bumps in
the autocorrelation, indicating the return  of sound waves through the
periodic boundary conditions of the  simulation box.  To minimize this
effect, it is necessary to use  a larger simulation box than typically
employed in molecular dynamics simulations of this kind.  We use a box
of 120 \AA{} and production runs of 20 ns.
\subsection{CG simulations}
To solve  numerically the SDE  given by Eq.~{(\ref{SDE-Model})} we have  upgraded the
G-J/F DPD algorithm proposed in  \cite{farago2016a} by including a new
Stokes  friction with  the corresponding  noise, as  described in  the
SM Section~S4.   The temperature of  the SDE, as well  as the
masses and charges  of the beads correspond to the  values observed in
the  AA simulation.  The length  of the  simulated CG  trajectories is
100 ns.

\subsection{Parameter estimation and validation}
The    set     of    parameters     of    the    model     given    by
Eqs.~{(\ref{SDE-Model})}-{(\ref{Fin:Noises})}  are noted  as
$\lambda=\left(\kappa_{\mu\nu},\omega_{\mu\nu}\right)$             and
$\gamma=\left(\gamma^S,\gamma^\bot,\gamma^{||}\right)$ where $\lambda$
are the static parameters and $\gamma$ the dynamic ones.  We propose a
conceptually   simple   self-averaging   method  to   evaluate   these
parameters.  Assume first  that  the dynamic  parameters $\gamma$  are
known. Then, we {\em enlarge} the  state space of the coarse variables
with  $\lambda$ and  propose  the following  coupled  dynamics in  the
enlarged space as,
\begin{eqnarray}
d a_t  &=&  F(a_t,\lambda_t,\gamma)dt
\label{aldyn0}\\
\dot{\lambda}_t &=&-\frac{1}{\tau}\left[ \llangle O\rrangle^{\rm mic} -O(a_t)\right]
\label{aldyn}
\end{eqnarray}
where $a_t=\left({{\bf  R}_{\mu},{\bf P}_{\mu}}\right)$  correspond to
the    CG   variables    and    $F(a_t,\lambda_t,\gamma$)    is   given    by
Eqs.~{(\ref{SDE-RP})}.  In Eq.~{(\ref{aldyn})} the functions
$O(a)$ represent  the selected \textit{target} quantities  (see below)
for  which   we  want   to  ensure   the  CG-AA   matching  conditions
$\llangle  O\rrangle^{\rm mic}  =  \llangle  O\rrangle^{\rm mes}$  for
their micro and mesoscopic averages, given by
\begin{align}
  \llangle O\rrangle^{\rm mic}
  &=\int dz\frac{e^{-\beta \hat{H}(z)}}{Z(\beta)} O(\hat{A}(z))
    \nonumber\\
    \llangle O\rrangle^{\rm mes}_\lambda
=&  \int d a     P^{\rm eq}(a)  O( a )
\end{align}
Here  $P^{\rm  eq}(a)$  is the  mesosocopic  equilibrium  distribution
Eq.~{(\ref{eq:11})}  and  the  microscopic  equilibrium  averages
$\llangle O\rrangle^{\rm  mic}$ can  be sampled from  {MD} simulations
and  constitutes  ``the  ground   truth''.   To  ensure  the  matching
condition,  the   parameters  $\lambda_t$  evolve  according   to  the
relaxation equation  Eq.~{(\ref{aldyn})}.  We assume that  the CG
dynamics in Eq.~{(\ref{aldyn0})} allow an ergodic sampling of the
CG phase space by the fast  variables $a$.  In this relatively general
scenario,   ergodicity  ensures   that  at   long  times   the  slower
$\lambda$-dynamics  in Eq.~{(\ref{aldyn})}  converges to  a fixed
point   where   the   CG   average  matches   the   AA   value,   i.e.
$\llangle    O\rrangle^{\rm    mes}=\llangle   O\rrangle^{\rm    mic}$
\cite{kiefer2001,givon2004}.       Thus,      at      long      times,
$\lambda(a_t) \rightarrow  \lambda^*$ which  determines the  sought CG
parameters.  This strategy allows one  to obtain the static parameters
entering  the  potential  of  mean  force  only,  but  can  be  easily
generalized to  dynamic transport parameters $\gamma$  coefficients by
considering   correlation   functions.   In  particular,   we   couple
Eqs.~(\ref{aldyn0}),(\ref{aldyn}) with
\begin{equation}
  \label{gamma}
\dot{\gamma}_t = -\frac{1}{\tau}\left[\langle \hat{Q}\hat{Q}(\tau^*) \rangle^{mic} - Q_t Q_{t-\tau^*}\right]
\end{equation}
where   $\langle   \hat{Q}\hat{Q}(\tau^*)    \rangle^{mic}$   is   the
equilibrium time correlation function of the observable $\hat{Q}(z_t)$
at time $\tau^*$,  measured in the MD simulation,  and $Q_t=Q(a_t)$ is
the value of the observable $Q$ obtained from Eq.~(\ref{aldyn0}).

Let us now  specify the observables $O(a),Q(a)$ that will  be used for
the  self-averaging process.  Focusing on  the static  parameters, the
observable  \( O(a)  \) is  inspired  by the  relative entropy  method \cite{shell2008}

(see Eq. (12) of \cite{shell2008})

\begin{align}
    \llangle\frac{\partial V^{\rm MF}(\hat{A},\lambda)}{\partial \lambda}\rrangle^{\rm mic}
  &=\llangle \frac{\partial V^{\rm MF}(a,\lambda)}{\partial \lambda}\rrangle^{\rm mes}_\lambda
    \label{RelEns-cond1}
\end{align}where the potential of
mean force in this model is given in Eq.~{(\ref{eq:47})}.
Therefore, Eq.~{(\ref{RelEns-cond1})} take the form
\begin{align}
  \label{eq:48}
  \llangle (\hat{R}_{\mu\nu}-\overline{R}_{\mu\nu})^2\rrangle^{\rm mic}
   &=\llangle (R_{\mu\nu}-\overline{R}_{\mu\nu})^2\rrangle^{\rm mes}
	\nonumber\\
\llangle \frac{1}{\hat{R}_{\mu\nu}}\rrangle^{\rm mic}
  &=\llangle  \frac{1}{R_{\mu\nu}}\rrangle^{\rm mes}
\end{align}
where the averages are
\begin{align}
  \label{eq:52}
  \llangle(\hat{R}_{\mu\nu}-\overline{R}_{\mu\nu})^2 \rrangle^{\rm mic}
  &=\int dz \rho^{\rm eq}(z) (\hat{R}_{\mu\nu}(z)-\overline{R}_{\mu\nu})^2
    \nonumber\\
  \llangle ({R}_{\mu\nu}-\overline{R}_{\mu\nu})^2\rrangle^{\rm mes}
  &=\int dR P^{\rm mes}(R) ({R}_{\mu\nu}-\overline{R}_{\mu\nu})^2
  \nonumber\\
\llangle \frac{1}{\hat{R}_{\mu\nu}}\rrangle^{\rm mic}
  &=\int dz \rho^{\rm eq}(z) \frac{1}{\hat{R}_{\mu\nu}(z)}
    \nonumber\\
  \llangle  \frac{1}{R_{\mu\nu}}\rrangle^{\rm mes}
  &=\int dR P^{\rm mes}(R) \frac{1}{R_{\mu\nu}}
\end{align}
and $\overline{R}_{\mu\nu}=\llangle |\hat{\bf R}_{\mu\nu}|\rrangle^{\rm mic}$ is  measured
from MD simulations. To  solve  Eqs.~{(\ref{eq:48})}  with the  self-averaging  method,  we
couple the SDE {(\ref{SDE-Model})} with  the following equations for the
parameters
\begin{align}
  \label{eq:StaticParameters}
  \frac{d\kappa_{\mu\nu}}{dt}
              &=-\frac{1}{\tau}\left(\llangle(\hat{R}_{\mu\nu}-\overline{R}_{\mu\nu})^2 \rrangle^{\rm mic}-(R_{\mu\nu}(t)-\overline{R}_{\mu\nu})^2\right)
              \nonumber\\
  \frac{d\omega_{\mu\nu}}{dt}
              &=-\frac{1}{\tau}\left(\llangle\frac{1}{\hat{R}_{\mu\nu}}\rrangle^{\rm mic}-\frac{1}{R_{\mu\nu}(t)}\right)
\end{align}
 In order to estimate the  dynamic parameters, we
will  also couple  the SDE  Eq.~{(\ref{SDE-Model})}  with a  set of  dynamic
equations         for         the        friction         coefficients
${\gamma}^{||},{\gamma}^{\bot},\gamma^S$.  We need as many correlation
functions as friction  coefficients need to be determined.  It is just
natural to  correlate the ``velocities'' that  accompany each friction
coefficient to produce the corresponding  friction force. In this way,
we propose the following set of equations for the friction parameters
\begin{align}
  \frac{d {\gamma}^{||}}{dt}
  &=\frac{1}{\tau}\sum_\mu
    \left[\llangle {\bf V}^{||}_{0\mu}\esc  {\bf V}^{||}_{0\mu}(\tau_{||}^*)\rrangle^{\rm mic}
    - {\bf V}^{||}_{0\mu}(t)\esc  {\bf V}^{||}_{0\mu}(t-\tau_{||}^*)\right]
    \nonumber\\
  \frac{d {\gamma}^{\bot}}{dt}
  &=\frac{1}{\tau}\sum_\mu
    \left[\llangle {\bf V}^{\bot}_{0\mu}\esc  {\bf V}^{\bot}_{0\mu}(\tau_{\bot}^*)\rrangle^{\rm mic}
    - {\bf V}^{\bot}_{0\mu}(t)\esc  {\bf V}^{\bot}_{0\mu}(t-\tau_{\bot}^*)\right]
    \nonumber\\
  \frac{{d \gamma}^S}{dt}
  &=\frac{1}{\tau} \left[\llangle {\bf V}_{\text{CoM}}\esc  {\bf V}_{\text{CoM}}(\tau_S^*)\rrangle^{\rm mic} - {\bf V}_{\text{CoM}}(t)\esc  {\bf V}_{\text{CoM}}(t-\tau_S^*)\right]
    \nonumber\\
  \label{eq:par1}
\end{align}
where we introduced the velocities,
\begin{align}
  {\bf V}^{\bot}_{0\mu}
  &\equiv \sum_\nu \left[{\bf 1}-{\bf e}_{\mu\nu}\esc {\bf e}_{\mu\nu}^T\right]\esc {\bf V}_{\mu\nu}
    \nonumber\\
  {\bf V}^{||}_{0\mu}
&\equiv   \sum_\nu    ({\bf V}_{\mu\nu}\esc{\bf e}_{\mu\nu}){\bf e}_{\mu\nu}
\end{align} and
the CoM velocity is ${\bf V}_{\text{CoM}} = \sum_{\mu} M_{\mu} {\bf V}_{\mu}/\sum_{\mu} M_{\mu}$.
The values of $\tau^*=\left\{\tau^*_{||},\tau_{\bot}^*,\tau_S^*\right\}$
were chosen as $\tau^*=\left\{150,120,250\right\} \text{fs}$.

\section{Results and Discussion}

The strategy to validate the proposed CG model  consists of three
phases, each involving different types of simulations:
\begin{enumerate}
\item  We first run MD  simulations  and  measure two  types  of averages:  the
  averages  $\llangle  \hat{O}  \rrangle^{\rm  mic}$  of  \emph{target
    observables} used in the self-averaging method and the averages of
  another  set of  \textit{validation observables}  used in  the third
  phase  below  to validate  the  CG  model.  The  target  observables
  $\hat{O}(z)$  are  provided  by  the  Relative  Entropy  Method,  as
  discussed  in Sec. \ref{Sec:REM},  Methods, and  include
  averages of  functions of the  distances between beads  appearing in
  Eq.~{(\ref{eq:52})},   in  particular   the  average   distance
  $\overline{R}_{\mu\nu}  \equiv  \llangle \hat{R}_{\mu\nu}  \rrangle^{\text
    mic}$,              the              inverse              distance
  $\llangle \frac{1}{\hat{R}_{\mu\nu}}\rrangle^{\rm mic}$ and the covariance
  matrix
  $\Sigma_{\mu\nu}\equiv\llangle(R_{\mu\nu}-\overline{R}_{\mu\nu})^2\rrangle^{\rm
    mic}$.       Moreover,      as      explained      in      Methods
  (Eqs.~{(\ref{eq:par1})}),  we include  dynamic information  and
  enforce  matching microscopic  time  correlations  of relative  bead
  velocities  and  CoM  velocity   \emph{at  a  particular  time  lag}
  $\tau^*$.  The CG model present  small variations with the values of
  $\tau^*$  selected for  fitting,  which provides  a  measure of  the
  quality of the CG model assumptions.

  Over the  MD runs one  also needs to measure  validation observables
  used to compare  the resulting CG model with the  AA model.  Here we
  consider  the  radial  distribution function (RDF) and  the beads'
  velocity autocorrelation function (VACF) \emph{at all times}.

\item Then we run the self-averaging method where Eq.~{(\ref{aldyn0})}
  is coupled to  Eqs.~{(\ref{aldyn})},{(\ref{gamma})} to determine the
  parameters of the model by  requiring that some selected observables
  have the same  mesoscopic and microscopic average.   Run the process
  until   convergence    to   the    fixed   point,    ensuring   that
  $\llangle O\rrangle^{\rm mes}  = \llangle \hat{O}\rrangle^{\rm mic}$
  and                                                               is
  $\llangle      QQ(\tau^*)\rrangle^{\rm      mes}     =      \llangle
  \hat{Q}\hat{Q}(\tau^*)\rrangle^{\rm mic}$  are satisfied,  to obtain
  the final  converged parameters $\lambda^*,\gamma^*$.  The evolution
  of  the   different  parameters  during  the   estimation  phase  is
  illustrated   in  Fig.    \ref{Fig:2a}.    It  can   be  seen   that
  $\lim_{t\rightarrow  \infty}\lambda_t   \rightarrow  \lambda^*$  and
  $\lim_{t\rightarrow  \infty}\gamma_t  \rightarrow \gamma^*$  with  a
  roughly exponential convergence at a rate given by the relaxation of
  the target observables $O(a)$.
\begin{figure}[h]
  \includegraphics[width=0.48\textwidth]{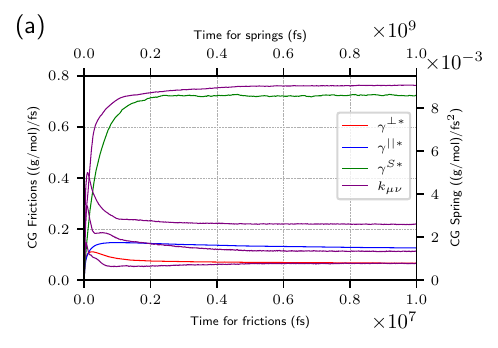}

\caption{The evolution of the CG parameters during the estimation
  phase,    reaching     convergence    to    the     sought    values
  ($\lambda,\gamma   \rightarrow  \lambda^*,\gamma^*$).}
  \label{Fig:2a}
\end{figure}

\item In the  final validation step, we  run Eq.~{(\ref{aldyn0})} with
  $\lambda=\lambda^*$,  $\gamma=\gamma^*$ and  compare the  CG and  AA
  validation observables (i.e. the RDF and VACF at all times).
\end{enumerate}

Applying  the above  protocol to  the  study of  the globular  protein
dynamics we  obtain a  range of elastic  constants $\kappa_{\mu\nu}^*$
from $(1.01\pm  0.67)\cdot 10^{-5}$  to $(1.56\pm  0.13)\cdot 10^{-2}$
with  an  average  of  $(1.94  \pm 0.15)\cdot  10^{-3}$  in  units  of
(g/mol)/fs$^2$.  Springs  with constants  smaller than  $10^{-5}$ have
been  set  to  zero.   In   Fig.   \ref{Fig:1}  we  present  a  visual
representation of  the beads joined  with a link whose  width reflects
the  strength of  the spring  constant $\kappa_{\mu\nu}$.   We observe
that the  spring constant generally  decreases with the  separation of
beads (see SM Section~S3).  The friction coefficients take
the      values      $\gamma^{S*}=0.72$,     $\gamma^{||*}=0.13      $
$\gamma^{\bot*}=0.07$ in units of (g/mol)/fs.
\begin{figure}[h]
    \includegraphics[width=0.5\textwidth]{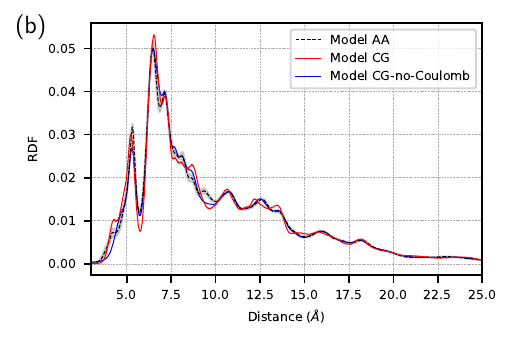}

 \caption{Radial distribution function  (RDF) of the beads  for the AA
   model (black) and different  mesoscopic models.  The standard error
   from 10 MD simulations of 20 ns is indicated by the gray area.  }
  \label{Fig:2b}
\end{figure}
Concerning the  model validation, Fig.  \ref{Fig:2b} shows the  RDF of
the  beads from  the microscopic  AA  and mesoscopic  CG models.   The
equilibrium   distribution   is   determined  by   the   elastic   and
electrostatic   reversible  terms   in  Eq.~{(\ref{eq:47})}.    Figure
\ref{Fig:2b} shows  the RDF of the  beads for both the  microscopic AA
(dashed black line) with the complete CG and the CG-no-Coulomb models,
the  later with  $\omega_{\mu\nu}=0$.  The  agreement between  both CG
models  and the  AA (black)  models is  remarkable.  Inclusion  of the
electrostatic terms leads to minor differences in the RDF, at least in
the presently  studied protein.   The first  peak of  the RDF  at 5.24
\AA{} gives the  typical distance between first  bead neighbors, which
compares       well       with        the       average       distance
$\overline{R}_{\mu\nu}=(5.32\pm 0.24)$ \AA, which  is the input in the
method.  The  second peak at  6.{54} \AA{} corresponds to  the typical
distance between  second neighbors  $(6.{48}\pm 0.{23})$  \AA{}.  Note
that the probability of finding a  bead at large distance from a given
one vanishes for distances larger than  the size of the protein (which
has  a typical  radius of  12.5 \AA,  according to  Fig \ref{Fig:2b});
hence, as opposed to the RDF  function of a liquid, the protein bead's
RDF vanishes at large distances.

\begin{figure}[h]
  \includegraphics[width=0.5\textwidth]{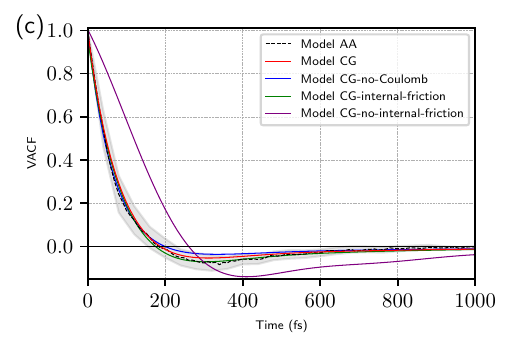}

\caption{Average
  VACF over  all beads for  the microscopic (black) and  for different
  mesoscopic models.  The gray area indicate the standard deviation of
  the VACF over different  beads.}
  \label{Fig:2c}
\end{figure}
On  the dynamic  side,  the  validation of  the  bead's VACF  requires
correct values  of both  elastic terms  and internal  and hydrodynamic
friction.  Each  bead presents  a slightly different  VACF due  to its
distinct connectivity. In Fig.  \ref{Fig:2c} we present the average
VACF  over all  beads and  indicate with  the gray  area the  standard
deviation  taken from  the  set  of individual  beads.  The error  bar
(standard error)  for the  individual bead VACF  is typically  3 times
smaller than  the standard deviation over  different beads, indicating
that the trends in individual bead VACF are significant and accurately
resolved.

We  recall (Methods  and Eqs.~{(\ref{eq:par1})})  that the  CG and  AA
VACFs  coincide  by construction  at  the  time-lag $\Delta  t=\tau^*$
selected in the parameter estimation step at which we enforce matching
with the AA time correlations.   Strictly, AA and CG time correlations
match at  two times, $\Delta t=\tau^*$  and at $\Delta t=0$  where the
later  is  guaranteed  by  the equipartition  theorem.   As  shown  in
Fig. \ref{Fig:2c}  this is  enough to  provide an  excellent agreement
with bead velocity time correlations (VACF) \textit{at all times}.  We
used $\tau^* \sim  200$ fs, which is about one-quarter  of the protein
main oscillation period, which can  be estimated by fitting the bead's
VACF in Fig.   \ref{Fig:2c} to $\cos(\omega t  +\phi) \exp(-\gamma t)$
(which  yields  $2\pi/\omega  \approx  900$ fs).   We  found  a  small
dependence  in the  parameter estimation  with the  selected $\tau^*$,
with  $5\%$ relative  variations in  the  CG parameters  in the  range
$\tau^* \in [50,  400]$ fs.  Such a dependence provides  a way to test
the CG  model assumptions, as improving  the CG model by  adding extra
physical  phenomena  and  transport  channels,  tends  to  reduce  its
dependence with $\tau^*$.

A very  useful physical insight can  be now obtained by  analyzing the
family           of            models           described           by
Eqs.~{(\ref{SDE-Model})}-{(\ref{Fin:Noises})}.   To    this   end   we
selectively  switch  off  their different  terms,  being  particularly
interested in analyzing the role of the internal friction.  The (full)
\textit{CG-model} takes into account all  the terms and parameters and
provides the  best comparison  between the  micro and  meso validation
average  observables.  The  \textit{CG-no-Coulomb} model  switches off
electrostatic   interactions   (setting   $\omega_{\mu\nu}=0$).    The
\textit{CG-no-Stokes} model switches off Stokes self-friction (setting
$\gamma^S=0$)  and \textit{CG-no-internal-friction}  switches off  the
internal  friction   only  (setting  $\gamma^{||}=0,\gamma^{\bot}=0$).
When  specific   parameters  are  set  to   zero,  the  self-averaging
estimation  process restarts,  allowing  for the  adjustment of  these
parameters as needed.

\begin{figure}[h]

    \includegraphics[width=0.5\textwidth]{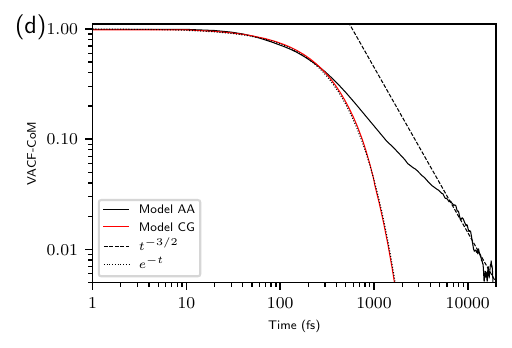}
 \caption{Protein center-of-mass velocity  time correlation (CoM-VACF)
   comparing the microscopic  (black) and CG model.   The log-log plot
   shows  two  main  dynamic  regimes:  short-time  exponential  decay
   $\exp[-\gamma t]$  and a long-time algebraic  decay $t^{-3/2}$. The
   long-time  hydrodynamic regime  originates  from  the diffusion  of
   solvent  vorticity, which  is not  represented as  a coarse-grained
   (CG) variable in  the DPD-Langevin model and,  therefore, cannot be
   captured by the current CG framework.  }
  \label{Fig:2d}
\end{figure}

Let us  start by  considering the CG-no-Stokes  model (green  lines in
Fig.   \ref{Fig:2c}) which  does  not includes  the self  hydrodynamic
friction.  As the internal  (blob-blob) forces conserve momentum, this
model predicts  that the  total momentum (i.e.   the CoM  momentum) is
strictly conserved  (we set it  to zero).   This obviously leads  to a
dramatic  failure  of the  CoM-VACF  which  vanishes by  construction.
Hence, the Stokes self-friction of the beads are certainly required to
describe the  protein diffusion.  In  this line,  note that if  we sum
Eq.~{(\ref{SDE-Model})} over all beads, all the forces satisfying
Newton's  Third  Law cancel  (in  particular  the reversible  and  the
internal friction forces). Hence, one is left with an equation for the
protein    center   of    mass    (CoM)   velocity    of   the    form
$d{\bf  V}_{\text{CoM}}=-\gamma\,{\bf  V}_{\text{CoM}}dt  +d\tilde{\bf
  V}_{\text{CoM}}$ which  predicts a  simple exponential decay  of the
CoM-VACF.   Introducing the  self-friction of  beads in  the CG  model
indeed leads an exponential decay  which perfectly matches the initial
decay  of  the microscopic  (AA)  CoM-VACF,  as  it should  [see  Fig.
\ref{Fig:2d}].
  To explain  the deviation  of the  CG model
CoM-VACF  at long  times,  we recall  that we  have  not included  the
vorticity diffusion  in the  CG description so  it cannot  capture the
algebraic decay at long-times which  is observed in the AA simulations
(Fig  \ref{Fig:2d}).  This is indeed the  celebrated long time
tail  $\sim  t^{-3/2}$  \cite{Alder1970,pomeau1975} arising  from  the
hydrodynamic propagation of momentum (vorticity).

A  surprising outcome  is  obtained when  analyzing the  CG-no-Coulomb
model, which  ignores the electrostatic interactions  (EI) (blue lines
in  Fig.    \ref{Fig:2b}  and  \ref{Fig:2c}).   While   switching  off
electrostatics  still  provides a  good  agreement  of the  RDF  [Fig.
\ref{Fig:2b}], it  leads to measurable  deviations in the  bead's VACF
[Fig. \ref{Fig:2c}], with a less  pronounced negative dip.  It is
  not entirely unreasonable  for electrostatics to have  a mild effect
  on the radial distribution  while significantly influencing velocity
  correlations.  The key  distinction lies  between ``where  the beads
  are'' and  ``how they move''.  The spatial arrangement of  beads may
  not be strongly  affected by electrostatics, whereas  their motion —
  captured through  velocity correlations —  can be influenced  by the
  long-range  nature  of   electrostatic  interactions.

Finally we set the internal friction coefficients to zero in the model
CG-no-internal-friction (violet lines), resulting  in a different bead
self-friction  coefficient.  As  shown  in  Fig. \ref{Fig:2c},  this
leads  to a  clearly  unacceptable disagreement  with the  microscopic
VACF.   It is  important to  state that  many CG  protein models  just
include self-friction, leading to  Langevin-based approaches which are
able to  recover the protein  equilibrium structure and the  root mean
squared (RMS) residues fluctuations,  but cannot reproduce the correct
vibrational relaxations of  the protein. Here, we conclude  that it is
necessary to introduce the \textit{internal} (DPD type) dissipation to
correctly model time-dependent internal protein fluctuation spectra.

\section{Conclusion}
In this paper, we present a  microscopically informed bead model for a
protein in water.  The model is  founded on the Mori-Zwanzig theory in
the Markovian  approximation \cite{hijon2010a} and  leads to a  set of
SDE for the  CG variables. The model describes the  interaction of the
beads with elastic and electrostatic  forces, and with friction forces
of  the DPD  and Langevin  type, describing  internal dissipation  and
solvent   dissipation,  respectively.    Both,   static  and   dynamic
parameters of the model have been obtained with a self-averaging
method that  couples the SDE with  a set of dynamic  equations for the
parameters,    which    evolve    over    a    significantly    longer
timescale.  Ergodicity  guarantees convergence  to  a  fixed point  in
parameter space, ensuring that the microscopic and mesoscopic averages
of selected observables are identical.

We  have  shown  that  by  comparing different  models  we  can  learn
important insights about how to describe appropriately the dynamics of
the protein  at a CG  level.  In particular,  we observe that  the DPD
friction between beads, representing \textit{internal dissipation}, is
numerically just as  significant as the solvent  dissipation caused by
Stokes friction.  This hightlights  that internal  dissipation is
  crucial  in  folded globular  proteins,  as  it is  in  unstructured
  unfolded proteins. Adding the Stokes self-friction of the beads, as
the minimal hydrodynamic effect, leads to a correct short-time protein
velocity  decorrelation,  which  however does  not  include  long-time
hydrodynamic tails. To capture this  effect, one should enlarge the CG
model to include transient vorticity transport (unsteady Stokes regime
\cite{Usabiaga2013}).   This can  be included  in the  CG model  using
Lagrangian            (fluctuating)           SDPD            solvents
\cite{Espanol2003SDPD,Ellero2018} or  immersed boundary  solvers based
on  continuum  fluctuating  hydrodynamics,  either  allowing  unsteady
vorticity  transport  \cite{Usabiaga2013}   or  adapted  to  Stokesian
hydrodynamics  \cite{Delong2014}.   In  any case,  these  descriptions
should be updated with an extra internal friction dissipation channel.
The  CG model  used hereby  does not  require to  include hydrodynamic
couplings between  different protein-beads,  confirming that  they are
strongly   screened   in   globlular  proteins.    Yet,   hydrodynamic
interactions  could be  relevant in  open chain  structures (e.g.   in
unstructured  proteins or  hinge  proteins  with multiple  equilibrium
states \cite{kapral2015}).  The present approach can be generalized to
add these and other transport phenomena in the CG description.

\section{Supplementary Material}
In S1 of the SM we discuss the existence of a proper separation of time scales needed to validate the proposed model. S2 gives details on the form of the random noise according to the Fluctuation-Dissipation Theorem. S3 plots the dependence between spring constants of the model and the distance between blobs. Finally, S4 gives the details on the integration algorithm for the DPD+Langevin CG equations.

\begin{acknowledgments}
We thank  Pablo Ibáñez  for his advice  on the use  of the  PDB.  This
research has been supported through MCIN grants PDC2021-121441-C22 and
PID2020-117080RB-C51, PID2020-117080RB-C54. We  acknowledge the computational  resources and
assistance provided by  the Centro de Computación  de Alto Rendimiento
CCAR-UNED.

\end{acknowledgments}

\end{document}

\title{Supplementary Material for ``Unravelling Internal Friction in a Coarse-Grained Protein Model''}

\author{Carlos Monago}
\affiliation{ Dept.   F\'{\i}sica Fundamental, Universidad Nacional
  de Educaci\'on a Distancia, Madrid, Spain}
\author{J.A. de la Torre}
\affiliation{ Dept.   F\'{\i}sica Fundamental, Universidad Nacional
  de Educaci\'on a Distancia, Madrid, Spain}
\author{R. Delgado-Buscalioni}
\affiliation{Dept.   F\'{\i}sica   de  la   Materia  Condensada,
  Universidad Autónoma de Madrid, Spain}
\author{ Pep Espa\~nol}
\affiliation{ Dept.   F\'{\i}sica Fundamental, Universidad Nacional
  de Educaci\'on a Distancia, Madrid, Spain}

\date{}
\maketitle
\setcounter{figure}{0}
\renewcommand{\thefigure}{S\arabic{figure}}
\setcounter{equation}{0}
\renewcommand{\theequation}{S\arabic{equation}}
\renewcommand{\thesection}{S\arabic{section}}

\section{The mapping}
\label{Sec:beads}
Many   different  ways   of  grouping   atoms,  i.e.    of  specifying
$\delta_\mu(i)$, in  a protein  have been devised.   As our  long term
interest is in the  elasto-mechanics properties of nanoscopic objects,
the mapping  that we consider  aims to minimize the  disparity between
the mass distribution of atoms and beads, and resembles a soft-Voronoi
partition of atoms into beads.

The idea  is to  minimize the  difference between  the microscopic
$ \hat{\rho}({\bf  r})$ and CG $\overline{\rho}({\bf r})$ density fields,
defined according to
\begin{align}
  \hat{\rho}({\bf r})&\equiv\sum_i^Nm_i\delta({\bf r}-{\bf r}_i)
                   \label{eq:1}\\
    \overline{\rho}({\bf r})&\equiv\sum_\mu^MM_\mu \Delta({\bf r}-{\bf R}_\mu) \label{eq:2}
\end{align}
 $m_i$ is
the mass of atom $i$, $M_\mu$ is the mass of the bead $\mu$ and $\Delta({\bf r}-{\bf R}_\mu)$ is a Gaussian given by
\begin{align}
  \label{eq:4}
  \Delta({\bf r}-{\bf R}_\mu)=\frac{1}{Z}\exp\left\{-\frac{({\bf r}-{\bf R}_\mu)^2}{2\overline{\sigma}^2}\right\}
\end{align}
where  $Z$ ensures  that  the  Gaussian is  normalized  to unity,  and
$\overline{\sigma}$ is the  width of the Gaussian. By  dividing by the
total  mass of  the  protein,  we may  interprete  $\hat{\rho}_{\bf r}$  and
$\overline{\rho}_{\bf  r}$   as  probability  distributions   and  the
objective   is  to   minimize   the  discrepancy   between  these   two
distributions.  Therefore,  we  construct  the  Kullback-Leibler  (KL)
divergence between the two distribution functions, defined as
\begin{align}
  \label{KL}
  D(\rho|\overline{\rho})&=\int d{\bf r}\hat{\rho}({\bf r})\ln\frac{\hat{\rho}({\bf r})}{\overline{\rho}({\bf r})}
\end{align}
The KL divergence becomes a function  of the bead configuration and we
ask which  configuration gives the smallest  KL divergence. Therefore,
we minimize (\ref{KL})
\begin{align}
  \label{eq:15}
  \frac{\partial   D(\rho|\overline{\rho})}{\partial {\bf R}_\mu}
  &=-\int d{\bf r}\hat{\rho}({\bf r})\frac{\partial}{\partial {\bf R}_\mu}\ln\overline{\rho}({\bf r})
\end{align}
where   the   $\hat{\rho}\ln\hat{\rho}$   term,    being   independent  on   $R$,
dissapears.  Setting  to  zero   the  partial  derivatives  and  using
(\ref{eq:2}) for the CG density field gives the following condition
\begin{align}
  \label{eq:17}
  0=\int d{\bf r}\frac{\hat{\rho}({\bf r})}{\overline{\rho}({\bf r})}\exp\left\{-\frac{({\bf r}-{\bf R}_\mu)^2}{2\overline{\sigma}^2}\right\}({\bf r}-{\bf R}_\mu)
\end{align}
Inserting the form (\ref{eq:1}) for the microscopic density field, (\ref{eq:17}) becomes
\begin{align}
    \label{eq:36}
\sum_im_i     \frac{1}{\overline{\rho}({\bf r}_i)}\exp\left\{-\frac{({\bf r}_i-{\bf R}_\mu)^2}{2\overline{\sigma}^2}\right\}({\bf r}_i-{\bf R}_\mu)=0
\end{align}
that can be written as
\begin{align}
  \label{eq:37}
  {\bf R}_\mu&=\frac{\sum_im_i{\bf r}_i\delta_\mu({\bf r}_i)}{\sum_im_i\delta_\mu({\bf r}_i)}
\end{align}
where we have introduced the Shepard functions
\begin{align}
  \label{eq:38}
  \delta_\mu({\bf r})&=\frac{\exp\left\{-\frac{({\bf r}-{\bf R}_\mu)^2}{2\overline{\sigma}^2}\right\}}{\sum_\nu\exp\left\{-\frac{({\bf r}-{\bf R}_\nu)^2}{2\overline{\sigma}^2}\right\}}
\end{align}
that form a partition of unity
\begin{align}
  \label{eq:39}
  \sum_\mu\delta_\mu({\bf r})=1 ,\,\forall\,{\bf r}
\end{align}
These functions  depend on the  width of the  Gaussians and it  can be
shown  that  as  $\overline{\sigma}\to0$,  this  is,  for  very  sharp
Gaussians, the  Shepard functions tend to  the characteristic function
of the Voronoi cells centered around ${\bf R}_\mu$ \cite{Flekkoy1999}.

Equation  (\ref{eq:37}) is  a  non-linear equation  for ${\bf  R}_\mu$
because  the   Shepard  functions   depend  on  the   beads  positions
themselves. Therefore, the positions of  the beads will be obtained in
an iterative way, reminiscent of the Lloyd's algorithm for calculation
of centroidal tessellations \cite{Lloyd1982}. This is, we denote ${\bf R}^n_\mu$ as the
solution at the $n$-th iteration and write the iteration as
\begin{align}
  \label{eq:40}
    {\bf R}^{n+1}_\mu&=\frac{\sum_im_i{\bf r}_i\delta^n_\mu({\bf r}_i)}{\sum_im_i\delta^n_\mu({\bf r}_i)}
\end{align}
where $\delta^n_\mu({\bf r})$ is the Shepard function defined in terms
of the bead's positions ${\bf  R}^n_\mu$ at the $n$-th iteration. Once
the iterative  procedure has converged  we have the best  positions of
beads  that would  reproduce  the microscopic  density  (at the  scale
$\overline{\sigma}$). After convergence, the  mapping appearing in the
definition   of   the  CG   density field (\ref{eq:2})  is   provided  by
$\delta_\mu(i)=\delta_\mu({\bf  r}_i) $.   This partition  is computed
with  the  initial configuration  and  remains  constant in  the  time
evolution (this is,  the particles constituting a bead  are always the
same). The selected value of  $\sigma=0.053$ is sufficiently small for
$\delta_\mu(i)$ to be either 0 or 1.
\begin{figure}[t]
  \includegraphics[width=0.5\textwidth]{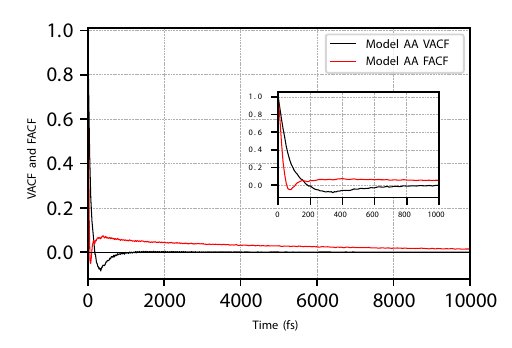}
  \caption{Comparison of the time-scales of the force autocorrelation function (FACF) and
    the velocity autocorrelation function (VACF). Inset is a zoom.}
  \label{Fig:1}
\end{figure}
\section{Separation of time scales}
The time scale of the  coarse-grained (CG) variables can be determined
using the  velocity autocorrelation function  (VACF) of the  beads. In
contrast,  the  memory time  scale  is  obtained from  the  Green-Kubo
formula, which  involves the correlation  of the forces acting  on the
beads. To  be precise,  the conditional  expectation of  the projected
forces appears in the Green-Kubo  formula. A reasonable proxy for such
a  conditional expectation  of the  projected forces,  that gives  the
memory  time scale,  is given  by the  force autocorrelation  function
(FACF).   As shown  in Fig.   \ref{Fig:1}, the  FACF displays  a rapid
initial  decay and  a  slow  subsequent decay.   As  it is  well-known
\cite{Espanol-Zuniga1993},  only the  rapid  decay  intervenes in  the
Green-Kubo expression and, therefore, it is what determines the memory
of the system.  In the bead  model presented in this paper, the memory
time  scale  differs  from  the  VACF   time  scale  by  a  factor  of
approximately four, as illustrated in Fig. \ref{Fig:1}

  \section{The    Fluctuation-Dissipation theorem}
\label{App:FD}
The                          noise                         covariances
$    d\tilde{\bf    P}^{\rm    b}_{\mu\mu'}    d\tilde{{\bf    P}^{\rm
    b}_{\nu\nu'}}^T$ are (to accomodate the Cartesian indices, we omit
the superindex ${\rm b}$)
  \begin{align}
    d\tilde{\bf  P}^{\alpha}_{\mu\mu'}          d\tilde{{\bf  P}}^{\beta}_{\nu\nu'}
    &=\left[A_{\mu\mu'}{\bf e}^\alpha_{\mu\mu'} dW_{\mu\mu'}+B_{\mu\mu'}\epsilon^{\alpha\alpha'\gamma}{\bf e}^{\alpha'}_{\mu\mu'}d{\bf V}^\gamma_{\mu\mu'}\right]\left[A_{\nu\nu'}{\bf e}^\beta_{\nu\nu'} dW_{\nu\nu'}+B_{\nu\nu'}\epsilon^{\beta\beta'\gamma'}{\bf e}^{\beta'}_{\nu\nu'} d{\bf V}^{\gamma'}_{\nu\nu'}\right]
      \nonumber\\
    &=A_{\mu\mu'}A_{\nu\nu'}{\bf e}^{\alpha}_{\mu\mu'} {\bf e}^\beta_{\nu\nu'} dW_{\mu\mu'}dW_{\nu\nu'}
      +B_{\nu\nu'}B_{\mu\mu'}\epsilon^{\alpha\alpha'\gamma}\epsilon^{\beta\beta'\gamma'}{\bf e}^{\beta'}_{\nu\nu'}{\bf e}^{\alpha'}_{\mu\mu'}     d{\bf V}^{\gamma}_{\nu\nu'}d{\bf V}^{\gamma'}_{\mu\mu'}
      \nonumber\\
    &=A_{\mu\mu'}A_{\nu\nu'}{\bf e}^\alpha_{\mu\mu'} {\bf e}_{\nu\nu'}^\beta \left[\delta_{\mu\nu}\delta_{\mu'\nu'}+\delta_{\mu\nu'}\delta_{\nu\mu'}\right]dt
      +B_{\nu\nu'}B_{\mu\mu'}
\underbrace{\epsilon^{\alpha\alpha'\gamma}\epsilon^{\beta\beta'\gamma}}_{\delta^{\alpha\beta}\delta^{\alpha'\beta'}-\delta^{\alpha\beta'}\delta^{\alpha'\beta}}{\bf e}^{\beta'}_{\nu\nu'}{\bf e}^{\alpha'}_{\mu\mu'}
\left[\delta_{\mu\nu}\delta_{\mu'\nu'}+\delta_{\mu\nu'}\delta_{\nu\mu'}\right]dt
      \nonumber\\
    &=A_{\mu\mu'}A_{\nu\nu'}{\bf e}^\alpha_{\mu\mu'} {\bf e}_{\nu\nu'}^\beta \left[\delta_{\mu\nu}\delta_{\mu'\nu'}+\delta_{\mu\nu'}\delta_{\nu\mu'}\right]dt
      +B_{\nu\nu'}B_{\mu\mu'}\left[\delta^{\alpha\beta}{\bf e}^{\alpha'}_{\nu\nu'}{\bf e}^{\alpha'}_{\mu\mu'}
-{\bf e}^{\alpha}_{\nu\nu'}{\bf e}^{\beta}_{\mu\mu'}\right]
\left[\delta_{\mu\nu}\delta_{\mu'\nu'}+\delta_{\mu\nu'}\delta_{\nu\mu'}\right]dt
  \end{align}
  Summing over $\mu',\nu'$
  \begin{align}
    d\tilde{\bf  P}^{\alpha}_{\mu}          d\tilde{{\bf  P}}^{\beta}_{\nu}
    &=\sum_{\mu'\nu'}A_{\mu\mu'}A_{\nu\nu'}{\bf e}^\alpha_{\mu\mu'} {\bf e}_{\nu\nu'}^\beta \left[\delta_{\mu\nu}\delta_{\mu'\nu'}+\delta_{\mu\nu'}\delta_{\nu\mu'}\right]dt
      \nonumber\\
    &+\sum_{\mu'\nu'}B_{\nu\nu'}B_{\mu\mu'}\left[\delta^{\alpha\beta}{\bf e}^{\alpha'}_{\nu\nu'}{\bf e}^{\alpha'}_{\mu\mu'}
-{\bf e}^{\alpha}_{\nu\nu'}{\bf e}^{\beta}_{\mu\mu'}\right]
\left[\delta_{\mu\nu}\delta_{\mu'\nu'}+\delta_{\mu\nu'}\delta_{\nu\mu'}\right]dt
      \nonumber\\
    &=\sum_{\mu'\nu'}A_{\mu\mu'}A_{\nu\nu'}{\bf e}^\alpha_{\mu\mu'} {\bf e}_{\nu\nu'}^\beta \delta_{\mu\nu}\delta_{\mu'\nu'}dt
+      \sum_{\mu'\nu'}A_{\mu\mu'}A_{\nu\nu'}{\bf e}^\alpha_{\mu\mu'} {\bf e}_{\nu\nu'}^\beta \delta_{\mu\nu'}\delta_{\nu\mu'}dt
      \nonumber\\
    &+\sum_{\mu'\nu'}B_{\nu\nu'}B_{\mu\mu'}\left[\delta^{\alpha\beta}{\bf e}^{\alpha'}_{\nu\nu'}{\bf e}^{\alpha'}_{\mu\mu'}
-{\bf e}^{\alpha}_{\nu\nu'}{\bf e}^{\beta}_{\mu\mu'}\right]
      \delta_{\mu\nu}\delta_{\mu'\nu'}dt
      +\sum_{\mu'\nu'}B_{\nu\nu'}B_{\mu\mu'}\left[\delta^{\alpha\beta}{\bf e}^{\alpha'}_{\nu\nu'}{\bf e}^{\alpha'}_{\mu\mu'}
-{\bf e}^{\alpha}_{\nu\nu'}{\bf e}^{\beta}_{\mu\mu'}\right]
\delta_{\mu\nu'}\delta_{\nu\mu'}dt
      \nonumber\\
    &=\delta_{\mu\nu}\sum_{\mu'}A_{\mu\mu'}A_{\mu\mu'}{\bf e}^\alpha_{\mu\mu'} {\bf e}_{\mu\mu'}^\beta dt
+ A_{\mu\nu}A_{\nu\mu}{\bf e}^\alpha_{\mu\nu} {\bf e}_{\nu\mu}^\beta dt
      \nonumber\\
    &+\delta_{\mu\nu}\sum_{\mu'}B_{\mu\mu'}B_{\mu\mu'}
      \left[\delta^{\alpha\beta}{\bf e}^{\alpha'}_{\mu\mu'}{\bf e}^{\alpha'}_{\mu\mu'}
      -{\bf e}^{\alpha}_{\mu\mu'}{\bf e}^{\beta}_{\mu\mu'}\right]
dt
      +B_{\nu\mu}B_{\mu\nu}\left[\delta^{\alpha\beta}{\bf e}^{\alpha'}_{\nu\mu}{\bf e}^{\alpha'}_{\mu\nu}
-{\bf e}^{\alpha}_{\nu\mu}{\bf e}^{\beta}_{\mu\nu}\right]
dt
      \nonumber\\
    &=\delta_{\mu\nu}\sum_{\mu'}A^2_{\mu\mu'}{\bf e}^\alpha_{\mu\mu'} {\bf e}_{\mu\mu'}^\beta dt
-A^2_{\mu\nu}{\bf e}^\alpha_{\mu\nu} {\bf e}_{\mu\nu}^\beta dt
      \nonumber\\
    &+\delta_{\mu\nu}\sum_{\mu'}B^2_{\mu\mu'}
      \left[\delta^{\alpha\beta}
      -{\bf e}^{\alpha}_{\mu\mu'}{\bf e}^{\beta}_{\mu\mu'}\right]
dt
      -B^2_{\mu\nu}\left[\delta^{\alpha\beta}
-{\bf e}^{\alpha}_{\mu\nu}{\bf e}^{\beta}_{\mu\nu}\right]
dt
  \end{align}
  According  to  the Fluctuation-Dissipation Theorem (FDT) (15) in the main manuscript (MMS),  with  the model (26,27(MMS)) for the dissipative matrix, when  $\mu \neq \nu$ we conclude
  that the noise amplitudes are given in terms of the friccion coefficients as
  \begin{align}
  A_{\mu\nu}^2
  &=2k_BT\gamma_{\mu\nu}^{\rm bb||}
&&\quad
B_{\mu\nu}^2  =2k_BT\gamma^{\bot}
\end{align}

In summary, the noise (31(MMS)) takes the form
\begin{align}
      d\tilde{\bf  P}^{\rm  b}_{\mu\nu}
  &=\sqrt{2k_BT\gamma_{\mu\nu}^{\rm bb||}}
    {\bf e}_{\mu\nu} dW_{\mu\nu}
   +\sqrt{2k_BT\gamma^{\bot}}{\bf e}_{\mu\nu}\times d{\bf V}_{\mu\nu}
\end{align}

\newpage

\section{Spring constants versus distances}
\begin{figure}[t]
  \includegraphics[width=0.45\textwidth]{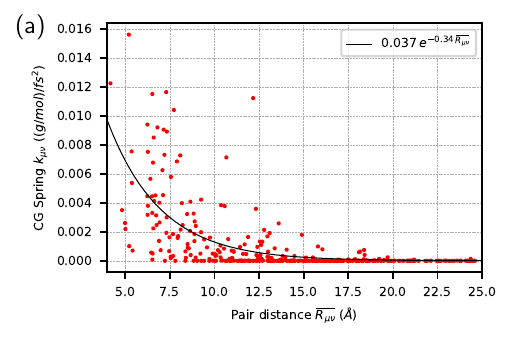}  \includegraphics[width=0.45\textwidth]{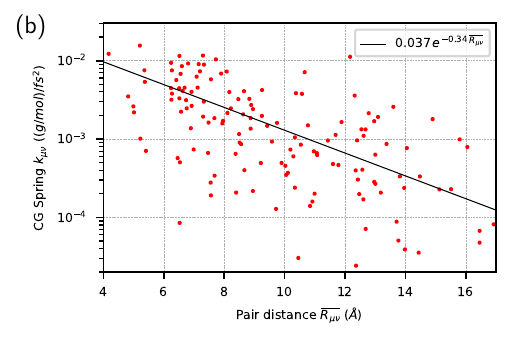}
  \caption{Spring constant as a function of pair average distance. (a) lin-lin plot. (b) log-lin plot.}
  \label{Fig:spring}
\end{figure}

The spring constant  $\kappa_{\mu\nu}$ of the spring  between the pair
$\mu,\nu$ is plotted  in Fig. \ref{Fig:spring} ((a) in  a linear scale
and  (b) in  a  semilogarithm  scale) as  a  function  of the  average
distance  $\overline{R}_{\mu\nu}$ between  the pair.  We observe  that
closer beads have generally larger spring constants.

\section{ The integrator}
Let us denote in a compact notation
\begin{align}
  R&=\{{\bf  R}_\mu,\mu=1,\cdots,M\}
     \nonumber\\
  V&=\{{\bf  F}_\mu,\mu=1,\cdots,M\}
\end{align}
We write the SDE governing these variables as
\begin{align}
  dR &= V dt
       \nonumber\\
  \mathbb{M} dV& = F^{\rm rev}(R)dt+F^{\rm irr}(R,V)dt+F^{\rm ran}(R)dt
\end{align}
and  $\mathbb{M}$ is  a diagonal  mass matrix  with elements  $m_\mu$,
while $F^{\rm  rev},F^{\rm irr},F^{\rm ran}$ stand  by the reversible,
irreversible, and random forces.

The integrator  updates  from the  values  $R_0,V_0$ at  the
beginning of  the time step  to the values $R_1,V_1$  at the end  of the
time step, which has duration $\Delta t$. The algorithm is as follows

\begin{enumerate}
  \item {\bf Initial conservative and dissipative forces:}
    \begin{align}
      F^{\rm rev}_0&\equiv F^{\rm rev}(R_0)
             \nonumber\\
      F^{\rm irr}_0&\equiv F^{\rm irr}(R_0,V_0)
    \end{align}

\item {\bf Positions with a deterministic half step:}
  \begin{align}
    R_{1/2}\equiv R_0+V_0\frac{\Delta t}{2}+\mathbb{M}^{-1}(F^{\rm rev}_0+F^{\rm irr}_0)\left(\frac{\Delta t}{2}\right)^2
  \end{align}

\item {\bf Random force at half step:}
  \begin{align}
    F^{\rm ran}_{1/2}\equiv F^{\rm ran}(R_{1/2})
  \end{align}

\item {\bf Velocity at half step using the previous random force:}
  \begin{align}
    V_{1/2}&\equiv V_0+\mathbb{M}^{-1}(F^{\rm rev}_0+F^{\rm irr}_0+F^{\rm ran}_{1/2})\left(\frac{\Delta t}{2}\right)
  \end{align}

\item {\bf Dissipative force at half step:}
  \begin{align}
    F^{\rm irr}_{1/2}&\equiv F^{\rm irr}(R_{1/2},V_{1/2})
  \end{align}

\item {\bf Position at full step:}
  \begin{align}
    R_1&\equiv R_0+V_0\Delta t+\mathbb{M}^{-1}(F^{\rm rev}_0+F^{\rm irr}_{1/2}+F^{\rm ran}_{1/2})\frac{\Delta t^2}{2}
  \end{align}

\item {\bf Reversible force at full step:}
  \begin{align}
    F^{\rm rev}_1&\equiv F^{\rm rev}(R_1)
  \end{align}

\item {\bf Recompute velocity at half step:}
  \begin{align}
    W_{1/2}&\equiv \frac{R_1-R_0}{\Delta t}
  \end{align}

\item {\bf Recompute irreversible force at half step:}
  \begin{align}
    \overline{F}_{1/2}^{\rm irr}\equiv F^{\rm irr}(R_{1/2},W_{1/2})
  \end{align}

\item {\bf Velocity at full step:}
  \begin{align}
    V_1&\equiv V_0+\mathbb{M}^{-1}\left(\frac{F^{\rm rev}_0+F_1^{\rm rev}}{2}+\overline{F}^{\rm irr}_{1/2}+F^{\rm ran}_{1/2}\right)\Delta t
  \end{align}

\item {\bf Irreversible force at full step:}
  \begin{align}
    F^{\rm irr}_1&\equiv F^{\rm irr}(R_1,V_1)
  \end{align}
\item {\bf Update}
  \begin{align}
    R_0&=R_1
         \nonumber\\
    V_0&=V_1
         \nonumber\\
    F^{\rm rev}_0
       &=F^{\rm rev}_1
         \nonumber\\
    F^{\rm irr}_0&=F^{\rm irr}_1
  \end{align}
\item
  {\bf Repeat} from step 2.
\end{enumerate}
We need to compute $F^{\rm ran},F^{\rm rev}$ once per time step, and $F^{\rm irr}$ twice per time step.

\newpage
\section{The Relative Entropy method}
\label{ME}
Shell  proposed    a  method  to  obtain  parametrised  models $P^{\rm mes}(a,\lambda)$ that
approximate  $ P^{\rm mic}(a)$ \cite{Shell2008a}.  The  essence of
Shell proposal is the  realisation that the Kullback-Leibler divergence
or relative entropy functional
\begin{align}
D_{\rm KL}(\lambda)
  &= \int da P^{\rm mic}(a)\ln\frac{P^{\rm mic}(a)}{P^{\rm mes}(a,\lambda)},
    \label{KL}
\end{align}
has  a  unique  global  minimum  when  the  {\em  model}  distribution
$P^{\rm mes}(a,\lambda)$,  is equal  to the {\em  target} distribution
$P^{\rm  mic}(a)$.   Therefore,  the   minimisation  with  respect  to
$\lambda$  leads to  an  approximant for  $P^{\rm mic}(a)$.   However,
$P^{\rm mic}(a)$ is unknown -- it is precisely what we want to know in
the first place. Nevertheless,  by noting the microscopic relationship
\begin{align}
  P^{\rm mic}(a)=\int dz\hat{\rho}^{\rm mic}(z)\delta\hat{A}(z)-a)
\end{align}
 we may write Eq. (\ref{KL}) as
\begin{eqnarray}
D_{\rm KL}(\lambda) &=&-\int dz \rho^{\rm mic}(z)\ln P^{\rm mes}(\hat{A}(z),\lambda)
+S_0\label{S1}
\end{eqnarray}
The  term  $S_0$ is  closely  related  to  the {\em  mapping  entropy}
introduced by Shell and plays no role in the following because it is a
term independent  of $\lambda$.   This function  has a  global minimum
that  occurs for  that $\lambda^*$  that gives  the model,  within the
family  of   parametrised  models,   closer  to  the   target  $P^{\rm
  eq}(a)$.    Note   that    the   evaluations    of   the    function
$D_{\rm KL}(\lambda)$ (or its  derivatives) simply involve equilibrium
averages that may  be computed with molecular dynamics  or Monte Carlo
simulations.  Therefore,  this is  an operative recipe  for optimising
coarse-grained models by using the  microscopic dynamics of the system
\cite{Shell2008a}.

In terms of an dimensionless free energy the probability distribution takes
the form
\begin{align}
  \label{eq:319}
  P^{\rm mes}(a,\lambda)=\frac{e^{-{\calF}(a,\lambda)}}{\int da' e^{-{\calF}(a',\lambda)}}
\end{align}
Substitution in (\ref{S1}) gives
\begin{align}
  \label{eq:320}
D_{\rm KL}(\lambda) &=\llangle {\calF}(\hat{A},\lambda)\rrangle^{\rm mic}+\ln \int da e^{-{\calF}^{\rm mes}(a,\lambda)}
\end{align}
Minimization with respect to $\lambda$ gives the condition
\begin{align}
  \label{eq:321}
  \llangle\frac{\partial {\calF}(\hat{A},\lambda)}{\partial \lambda}\rrangle^{\rm mic}
  &=\llangle \frac{\partial {\calF}(a,\lambda)}{\partial \lambda}\rrangle^\lambda
\end{align}
This is, the derivatives with respect to the parameters have the same micro and macro averages.
In  explicit terms
\begin{align}
  \label{eq:322}
  \int dz \rho^{\rm mic}(z)\frac{\partial  {\calF}(\hat{A}(z),\lambda)}{\partial \lambda}
  &=\int da P^{\rm mes}(a,\lambda)\frac{\partial  {\calF}(a,\lambda)}{\partial \lambda}
\end{align}
When  dealing  with a  free  energy  model  which  is linear  in  the
parameters,    the   right-hand    side    becomes   independent    of
$\lambda$. Thus, we only  require a single
molecular dynamics simulation, which serves as the reference or ground
truth.

\bibliography{Supp}